\documentclass[%
reprint,
superscriptaddress,
nofootinbib,
amsmath,
amssymb,
aps,
prd,
floatfix,
showkeys,
]{revtex4-2}

\usepackage{graphicx}
\usepackage{dcolumn}
\usepackage{bm}
\usepackage{subfigure}

\usepackage{epsfig}
\usepackage{mathtools}
\usepackage{booktabs}
\usepackage{aas_macros}

\usepackage{natbib}
\usepackage[colorlinks]{hyperref}
\hypersetup{linkcolor=magenta,citecolor=blue}
\usepackage[colorlinks]{hyperref}

\newcommand{\beq}{\begin{equation}}
\newcommand{\eeq}{\vspace{0cm} \end{equation}}
\newcommand{\beqq}{\setlength\arraycolsep{2pt}\begin{eqnarray}}
\newcommand{\eeqq}{\vspace{0cm} \end{eqnarray}}

\begin{document}

\title{Dark interactions in neutron star interiors: \\ the interplay of baryons, dark matter, and dark energy}

\author{L. F. Araújo}
\email{loreanyfa@usp.br}
\affiliation{Departamento de Astronomia, Universidade de S\~{a}o Paulo \\ Rua do Mat\~{a}o, 1226 - 05508-900, S\~ao Paulo, SP, Brazil}

\author{G. Lugones}
\email{german.lugones@ufabc.edu.br}
\affiliation{Universidade Federal do ABC, Centro de Ci\^encias Naturais e Humanas, Avenida dos Estados 5001- Bang\'u, CEP 09210-580, Santo Andr\'e, SP, Brazil.}

\author{J. A. S. Lima}
\email{jas.lima@iag.usp.br}
\affiliation{Departamento de Astronomia, Universidade de S\~{a}o Paulo \\ Rua do Mat\~{a}o, 1226 - 05508-900, S\~ao Paulo, SP, Brazil}

\begin{abstract}
The impact of energy exchange among the three relevant fluid components—baryonic matter, fermionic dark matter (DM), and vacuum-like dark energy (DE)—on the internal structure of neutron stars is investigated. Working with a representative DM mass $m_\chi=10~\mathrm{GeV}$ and a barotropic DE relation $p_{\mathrm{de}}=\omega \,\epsilon_{\mathrm{de}}$, we embed one or two phenomenological source terms $Q_i$ in the Tolman–Oppenheimer–Volkoff equations and analyse three hierarchical scenarios: (i) coexisting but non-interacting fluids (Model~I); (ii) fully interacting fluids with independent baryon–DM and DM–DE exchanges and optional DM self-repulsion (Model~II); and (iii) a unified dark sector that couples to baryons through a single exchange term while retaining an internal DM/DE partition (Model~III).  Across all models two complementary mechanisms dominate: softening by massive, pressure-poor DM, and vacuum softening/binding by the negative pressure of DE.  Model~I exhibits these effects in their purest form, yielding either softened or self-bound stars depending on the dark composition.  In Model~II the exchange terms self-regulate, cancelling the explicit dependence on the fluid coupling strength $\alpha$, so that mass–radius curves are nearly invariant when $\alpha$ varies from $0.05$ to $1000$.  Model~III breaks this $\alpha$-degeneracy: when the dark sector retains a substantial vacuum fraction, the baryonic pressure gradient is strongly suppressed and both the maximum mass and radii decrease appreciably, whereas a pure-DM core remains almost insensitive to the exchange strength.  The study delineates the conditions under which dark interactions can (or cannot) imprint macroscopic signatures, thereby providing a theoretical baseline for future, more strongly constrained, microphysical models.
\end{abstract}


\maketitle
\bigskip

\section{Introduction}

Observations across multiple astrophysical and cosmological scales have revealed that the vast majority of the Universe is composed of unknown forms of matter and energy, collectively referred to as dark matter (DM) and dark energy (DE). Standard cosmology explains many large-scale phenomena by positing that DM contributes to structure formation through gravitational clustering, while DE drives the observed late-time acceleration of the Universe \citep{perivolaropoulos2022challenges, wang2016dark, alcaniz2005interpreting, perico2013complete, lima2000radiation}. Yet, despite remarkable progress in cosmology, the fundamental nature of these dark components remains elusive.

The role of dark energy (DE) in high-density, small-scale environments such as neutron stars remains much less explored than its cosmological implications. Early theoretical work proposed exotic compact objects whose interiors were partly or wholly a vacuum-energy fluid. \citep{gliner1966algebraic,Bardeen1968,Chapline:2004jfp,Mazur:2001fv}, leading to concepts such as DE stars \citep{Chapline:2004jfp}. Indeed, already in 1966, \citet{gliner1966algebraic} suggested that negative vacuum pressure could impact particle balance via electrostatic repulsion, anticipating that a nontrivial vacuum component might arise in regions of high density, potentially decaying in ways that induce collapse. Contemporaneously, \citet{sakharov1966initial} argued that an equation of state $p = -\epsilon$ might be necessary for the formation of early cosmic structures. In later decades, various authors continued to explore how changing vacuum energy could arise from the conservation of matter left over by an inflationary period, often coupled with a time- or position-dependent $G$ \citep{sola2008dark} or with growing vacuum energy in gravitational collapse \citep{Campos_2012}. More recently, there has been a resurgence of interest in how cosmic vacuum energy could impact neutron stars directly, particularly in the form of objects possessing inhomogeneous vacuum interiors \citep{Dymnikova:1992ux,dymnikova2002cosmological,Campos_2012, das2023dark}, or with anisotropic pressure profiles \citep{posada2017slowly,mazur2023gravitational,novosyadlyj2014dynamics}.  In parallel, other works have considered how a DE fluid, possibly interacting with matter or combined with modified gravity, can affect the mass and radius of neutron and strange stars \citep{Astashenok:2023ewh}.

Modern numerical and semi-analytic studies also explore the possibility that vacuum energy changes in astrophysical phases, thus contributing non-negligibly to the structure of neutron stars. Indeed, the way vacuum energy might modify stellar equilibrium and gravitational observables has motivated investigations of new QCD phases and their corresponding vacuum expectation values inside neutron stars \citep{Bellazzini:2015wva, Csaki:2018fls}. Our own earlier work \citep{Araujo:2024txe} considered a two-component scenario in which standard nuclear matter coexists with a vacuum fluid obeying $p = -\epsilon$, radially varying from the stellar core outward. That study showed how a fraction of vacuum energy can alter neutron-star maximum masses and radii in ways consistent with observational bounds.

In contrast, neutron-star research has long considered the possible impact of DM on these compact objects \citep{Li:2012qf,Mukhopadhyay:2015xhs,Baym:2018ljz,Ellis:2018bkr,Dengler:2021qcq,Kain:2021hpk,Sen:2021wev,Jimenez:2021nmr,Miao:2022rqj,Mariani:2023wtv}. Such ``admixed’’ neutron stars allow for scenarios in which DM, ranging from condensate states to free or self-interacting fermions, coexists with ordinary nuclear matter. This coexistence may lead to DM cores \citep{Li:2012qf,Jimenez:2021nmr}, DM halos \citep{Miao:2022rqj}, or more smoothly distributed DM profiles \citep{Kain:2021hpk}, each producing distinct imprints on the mass-radius relation, tidal deformability, stability against radial oscillations, and other observable properties. In many of these scenarios, the precise properties of DM—whether it is strongly self-interacting \citep{Mukhopadhyay:2015xhs, Mariani:2023wtv}, feebly interacting \citep{Baym:2018ljz, Sen:2021wev}, or mirrorlike \citep{Kain:2021hpk}—can crucially determine how it modifies the star’s internal structure and gravitational footprint.

Despite these advances, the combined interplay among ordinary matter, DM, and DE within a single stellar object has seldom been studied in detail. Recent works have begun to address stars composed of both DM and DE \citep{Sepulveda:2024hdc,Jyothilakshmi:2024zqn}, examining radial and non-radial oscillations in exotic configurations. Still, the possible interactions that couple DM and DE—central to many cosmological models—could likewise manifest at the extreme densities and pressures of neutron-star interiors, thus affecting stellar formation, equilibrium conditions, and global observables. In principle, these \emph{unified} or \emph{coupled} dark-sector scenarios \citep{yoo2012theoretical} offer a promising way to address lingering cosmological puzzles such as the coincidence problem, while simultaneously providing a more complete picture of the physics of compact objects.

Here, we take a significant step by proposing a three-component framework in which neutron stars simultaneously host ordinary nuclear matter, DM, and DE. We further allow for the possibility that these three components can interact with each other, not only coexisting but exchanging energy or momentum. Our approach is motivated by the concept of coupled dark sectors in cosmology, wherein DM and DE interactions might alleviate fine-tuning issues on cosmic scales. Translated into the context of neutron stars, such interactions could leave distinctive imprints on maximum masses, radii, and tidal deformabilities in binary mergers, potentially becoming testable as observational precision continues to improve.

This paper is organized as follows. In Section~\ref{subsec:ordinary_matter_eos}, we describe the nuclear equations of state employed for ordinary matter in neutron stars, with a focus on piecewise polytropic representations of state-of-the-art models. In Section~\ref{subsec:DM_eos}, we then outline the equations of state for DM, considering both a degenerate fermion gas and possible self-interactions via a Yukawa-type force. Section~\ref{subsec:DE_eos} reviews a general formalism for DE, starting with the simplest $p = -\epsilon$ case and extending to more general parameterizations. By integrating these three components within a consistent Tolman-Oppenheimer-Volkoff (TOV) framework, we aim to explore how their mutual interactions affect the overall structure of neutron stars, thereby shedding light on fundamental physics both at nuclear densities and within the broader dark sector. To this end, we employ the general framework, including key parameters and boundary conditions, outlined in Sections~\ref{sec:general_setup} and~\ref{sec:closure_and_BC}. Our analysis considers three models: Model I (Section~\ref{sec:model_1}) captures the coupling among the three fluids through fractional relationships between their energy densities. Model II (Section~\ref{sec:model_2}) introduces interactions mediated by the dark matter component. Model III (Section~\ref{sec:model_3}) also incorporates these interactions, but considers the dark sector a unified component through fractions defined in terms of energy densities. The results of this analysis are presented in Section~\ref{sec:results}, and our conclusions are summarized in Section~\ref{sec:conclusions}.

Throughout our study, we make extensive use of modern observational constraints, including multi-messenger data from pulsar timing, X-ray measurements from the Neutron Star Interior Composition Explorer (NICER), and gravitational waves from neutron-star mergers detected by LIGO-Virgo. We present and discuss our results with special attention to how future measurements may help discriminate among different dark interaction models or constrain the fraction of dark components present in neutron stars.

\section{Equations of state}
\label{sec:EOS}

In this section, we outline the three distinct EOS that characterize the components of our stellar model: ordinary matter, DM, and DE. First, we describe how a piecewise polytropic approach can effectively model nuclear matter at high densities while preserving the essential microphysical properties of neutron-star interiors. Next, we consider the possibility that DM can be treated as a relativistic, degenerate Fermi gas, potentially modified by self-interactions if the particles are coupled via a Yukawa-type force. Finally, we turn to DE, where we adopt a general parameterization $p_{\mathrm{de}} = \omega \epsilon_{\mathrm{de}}$ to accommodate a range of cosmological scenarios, including the cosmological constant and dynamical scalar-field models. By combining these three EOSs, we aim to investigate how ordinary matter, DM, and DE could jointly influence the structure and properties of compact astrophysical objects.

\subsection{Equation of state for ordinary matter}
\label{subsec:ordinary_matter_eos}

One practical way to represent modern nuclear EOSs for dense matter is to approximate them with a segmented polytropic form. This piecewise polytropic approach offers a convenient framework for capturing the essential microphysical behavior of nuclear matter over different density ranges without resorting directly to highly complex functional forms. By adjusting the polytropic indices and the number of segments, one can reproduce a wide variety of nuclear EOSs with good accuracy.

In this work, we adopt the Generalized Piecewise Polytropic (GPP) representation introduced in Ref.~\cite{o2020parametrized}. An important feature of this scheme is that it enforces not only the continuity of pressure and energy density across segment boundaries but also the continuity of the speed of sound. Thus, the GPP model provides a smooth, thermodynamically consistent description of dense matter.

Within each polytropic segment, the pressure $p_{\mathrm{m}}$ and the energy density $\epsilon_{\mathrm{m}}$ are treated as functions of the rest-mass density $\rho_{\mathrm{m}}$. Specifically,
\begin{align}
p_{\mathrm{m}}(\epsilon_{\mathrm{m}}) &= K \rho_{\mathrm{m}}^{\Gamma} + A, \\
\epsilon_{\mathrm{m}}(\rho_{\mathrm{m}}) &= \frac{K}{\Gamma - 1} \,\rho_{\mathrm{m}}^{\Gamma}  + (1+a)\,\rho_{\mathrm{m}} - A,
\end{align}
from which one obtains the useful derivative
\begin{align}
\frac{d\epsilon_{\mathrm{m}}}{dp_{\mathrm{m}}} 
= \frac{1}{\Gamma - 1} 
+ \frac{(1+a)}{\Gamma\,K\,\rho_{\mathrm{m}}^{\Gamma - 1}}.
\end{align}
Here, $K$, $\Gamma$, $A$, and $a$ are coefficients that characterize the EOS in each segment.

To ensure smooth matching of adjacent segments at a chosen density $\rho_i$, continuity conditions for $p_{\mathrm{m}}$ and $\epsilon_{\mathrm{m}}$ (and their first derivatives) impose relationships among the coefficients in neighboring segments. For two segments labeled $i$ and $i+1$, these conditions require
\begin{align}
K_{i+1}  &= K_i \,\frac{\Gamma_i}{\Gamma_{i+1}} \,\rho_i^{\,\Gamma_i - \Gamma_{i+1}},  \\
A_{i+1} &= A_i  + \bigl(1 - \tfrac{\Gamma_i}{\Gamma_{i+1}}\bigr)\,K_i\,\rho_i^{\,\Gamma_i}, \\
a_{i+1}  &= a_i + \Gamma_i\,\frac{\Gamma_{i+1} - \Gamma_i}{(\Gamma_{i+1}-1)(\Gamma_i-1)}\, K_i\,\rho_i^{\,\Gamma_i - 1}.
\end{align}
These constraints guarantee that pressure, energy density, and speed of sound all behave smoothly when transitioning between segments.

From the wide range of EOSs that can be generated within the GPP framework, we focus on two models originally described in Ref.~\cite{o2020parametrized}, both of which are consistent with chiral effective field theory (EFT) approaches. Chiral EFT provides a systematic low-momentum expansion for nuclear forces and remains reliable up to densities a bit above the saturation density $n_0$ \cite{Coraggio:2012ca, Gandolfi:2011xu, Holt:2012yv, Hebeler:2009iv, Sammarruca:2012vb, Tews:2012fj}. 

\textbf{APR EOS.} Based on variational methods, the APR model \cite{Akmal:1998cf} uses nucleon-nucleon potentials and three-body forces fitted to scattering data and light nuclei properties. 

\textbf{MPA1 EOS.} In contrast, the MPA1 EOS \cite{Muther:1987xaa} stems from relativistic Brueckner-Hartree-Fock calculations, emphasizing tensor interactions mediated by pion and $\rho$ mesons. It yields a stiffer behavior at higher densities, favoring larger stellar radii and higher maximum masses.

Both the APR and MPA1 EOSs satisfy key observational constraints on neutron stars, such as the $\sim 2\,M_{\odot}$ mass measurements \cite{Demorest:2010ats, Antoniadis:2013amp, Cromartie:2020rsd, Fonseca:2021rfa} and recent radius determinations from NICER \cite{Miller:2019pjm,Riley2019anv, Miller:2021tro,Riley:2021anv}. By selecting two EOSs that differ in stiffness, we can examine how variations in the underlying microphysics influence neutron-star structure while maintaining consistency with experimental and observational data.

\subsection{Equation of state for DM}
\label{subsec:DM_eos}

The microscopic nature of DM is still unknown, yet cosmological and laboratory bounds allow a broad but finite mass window for weakly interacting massive particles (WIMPs), spanning roughly $\mathrm{MeV}$ to $\mathrm{PeV}$ scales \cite{abdalla2019brazilian}.  In this work we model DM as a spin-$1/2$ degenerate Fermi gas whose particles have a fixed mass
\begin{equation}
m_\chi = 10~\mathrm{GeV}.
\end{equation}
This value sits comfortably in the canonical WIMP range and, importantly, is large enough to avoid the formation of extended DM halos in neutron stars; halo–dominated configurations that arise for $m_\chi \sim 0.1-1 \mathrm{GeV}$ \cite{Miao:2022rqj} will be analysed in a separate publication.

\textit{Free Fermi gas.} Setting $\hbar=c=1$ and defining the dimensionless Fermi momentum $z\equiv k_{F\chi}/m_\chi$, the zero-temperature pressure and energy density of the \emph{free} gas read
\begin{align}
p_\chi^{\mathrm{FG}} &= 
  \frac{m_\chi^{4}}{24\pi^{2}}
  \Bigl[z\sqrt{z^{2}+1}\,(2z^{2}-3)+3\sinh^{-1}z\Bigr],\label{eq:FG_p}\\
\epsilon_\chi^{\mathrm{FG}} &=
  \frac{m_\chi^{4}}{8\pi^{2}}
  \Bigl[z\sqrt{z^{2}+1}\,(2z^{2}+1)-\sinh^{-1}z\Bigr].\label{eq:FG_e}
\end{align}
For $z\ll1$ (the regime relevant inside our stars) the leading terms reduce to 
\begin{equation}
p_\chi^{\mathrm{FG}}      \simeq \frac{m_\chi^4\,z^5}{15\pi^2},
\quad
\epsilon_\chi^{\mathrm{FG}} \simeq \frac{m_\chi^4\,z^3}{3\pi^2}.
\label{eq:FG_for_z_ll_1}
\end{equation}

\textit{Repulsive self-interactions.}   A short-range, vector-like repulsion of Yukawa range $m_{\mathrm int}^{-1}$ adds an identical
contribution to pressure and energy density,
\begin{equation}
\Delta p_\chi = \Delta\epsilon_\chi = \frac{m_\chi^{4}\,q^{2}z^{6}}{9\pi^{4}},\qquad   q\equiv\frac{m_\chi}{m_{\mathrm int}}.
\label{eq:vector_contribution_DM}
\end{equation}
Combining Eqs.\,\eqref{eq:FG_p}–\eqref{eq:FG_e} with $\Delta p_\chi=\Delta\epsilon_\chi$ yields the \emph{single} EOS we use throughout:
\begin{align}
p_\chi &= p_\chi^{\mathrm{FG}}+\Delta p_\chi,\\[3pt]
\epsilon_\chi &= \epsilon_\chi^{\mathrm{FG}}+\Delta\epsilon_\chi.
\label{eq:DM_EOS_full}
\end{align}
Two benchmark couplings are explored later: $q=0.1$ (\textit{weak}) and $q=1000$ (\textit{strong}) \cite{Miao:2022rqj}.

Including heavy, weakly–pressurised fermions is known to soften the composite EOS, decreasing both the maximum mass and radius of neutron stars \cite{panotopoulos2017dark,ellis2018dark,routaray2023probing}.  For $m_\chi\sim\mathrm{GeV}$ this softening typically pushes the mass–radius curve downward, yet still allows $M_{\max}\gtrsim2\,M_\odot$ \cite{li2012too, das2021dark, barbat2024comprehensive}.  Lighter particles ($m_\chi\lesssim0.1\,$GeV) combined with weak self-interaction ($q\lesssim0.1$) tend instead to form low-density halos that can increase the radius or maximum mass \cite{Miao:2022rqj}; those configurations are intentionally excluded here by our choice $m_\chi=10\,$GeV.

\subsection{Equation of state for dark energy}
\label{subsec:DE_eos}

Observations of baryon acoustic oscillations (BAO), Type Ia supernovae (SN Ia), and the cosmic microwave background (CMB) provide strong evidence that the DE equation of state is very close to
\begin{equation}
p_{\mathrm{de}} = -\,\epsilon_{\mathrm{de}},
\end{equation}
as in the case of a cosmological constant. In other words, $\omega \equiv p_{\mathrm{de}}/\epsilon_{\mathrm{de}} \approx -1$. However, this simple vacuum-energy model faces well-known theoretical challenges, such as the \emph{fine-tuning} or cosmological constant problem, in which the observed vacuum-energy density differs by about 120 orders of magnitude from naive quantum field theory estimates, as well as the \emph{coincidence problem}, which questions why the present-day energy densities of DM and DE are of the same order of magnitude.

To address these issues, various alternative models for DE have been proposed. \emph{Quintessence} and \emph{k-essence} models, for example, rely on a dynamical scalar field whose energy density drives the accelerated expansion of the Universe. Other proposals, such as \emph{unified or coupled dark sector} scenarios, posit direct interactions between DM and DE, aiming to alleviate the coincidence problem. Additionally, \emph{modified gravity} theories, including $f(R)$ gravity, incorporate higher-order curvature terms or introduce extra scalar degrees of freedom in Einstein’s equations \cite{yoo2012theoretical}, thereby offering further possibilities for explaining the observed DE phenomena.

A common phenomenological way to capture these scenarios is to adopt a more general parameterization for the DE EOS:
\begin{equation}
p_{\mathrm{de}} = \omega\,\epsilon_{\mathrm{de}},
\end{equation}
where $\omega$ may vary with time or scale factor. In many dynamical DE models, $\omega \approx -1$ in late cosmological epochs, recovering a behavior similar to the cosmological constant. From the Friedmann equations, it follows that an accelerated expansion requires $\omega < -\tfrac{1}{3}$ at late times. Consequently, models often impose a maximum value $\omega_{\max} = -\tfrac{1}{3}$ as the boundary between decelerating and accelerating phases of cosmic evolution (see, e.g., standard cosmology texts for detailed derivations).

\section{Stellar structure equations with dark components}
\label{sec:structure}

This section sets out the theoretical framework for describing neutron stars that contain both DM and DE. We begin in Sec.~\ref{sec:general_setup} with the general setup, introducing the spherically symmetric metric, the total energy--momentum tensor (including baryonic and dark components), and the resulting stellar structure equations. In Sec.~\ref{sec:closure_and_BC}, we discuss the closure requirements and boundary conditions needed to solve these equations.

Next, we present three distinct models for incorporating DM and DE:
\begin{itemize}
\item \emph{Model~I: coupled fluids} (Sec.~\ref{sec:model_1}). We assign fractions of the total energy density to each component, allowing these fractions to be either constant or radially varying. 

\item \emph{Model~II: interacting fluids with DM--mediated exchange} (Sec.~\ref{sec:model_2}). In this approach, explicit interaction terms $Q_{i}$ govern the local energy--momentum transfer among baryons, DM, and DE. DM is taken as the primary mediator, and each fluid’s pressure equation is modified accordingly.

\item \emph{Model~III: interacting fluids with a unified dark sector} (Sec.~\ref{sec:model_3}). Here, DM and DE are treated as a single fluid that exchanges energy only with baryons via a single interaction term. An additional fraction $x_{\chi}$ partitions the unified dark sector between matterlike and vacuumlike components, making this model a hybrid of the coupled and interacting approaches.
\end{itemize}
By comparing these formulations, we explore how different assumptions about the dark sector influence the stellar structure and mass--radius profiles of neutron stars.

\subsection{General setup}
\label{sec:general_setup}

To investigate the influence of dark components on neutron star structure, we model the star as a static, isotropic, and spherically symmetric system. The spacetime geometry of such a star is described by the metric
\begin{equation}
ds^2 = e^{2\nu(r)}\, dt^2 - e^{2\mu(r)}\, dr^2 - r^2 \bigl(d\theta^2 + \sin^2\theta\, d\phi^2\bigr),
\label{eq:metricsch}
\end{equation}
where we work in natural units ($c = G = 1$). For convenience, we define 
$
e^{2\mu(r)} \equiv 1 - \frac{2m(r)}{r},
$
where the function $m(r)$ represents the mass enclosed within radius $r$.

The neutron star is assumed to comprise three components: baryonic matter, DE, and DM. Each is treated as an independent perfect fluid, so that the total energy--momentum tensor can be written as
\begin{equation}
T_{\mu \nu} = T^\mathrm{m}_{\mu \nu} + T^{\mathrm{de}}_{\mu \nu} + T^{\chi}_{\mu \nu},
\end{equation}
where $T^\mathrm{m}_{\mu \nu}$, $T^{\mathrm{de}}_{\mu \nu}$, and $T^\chi_{\mu \nu}$ denote the respective contributions from baryonic matter, DE, and DM. For any perfect fluid, the energy--momentum tensor takes the form
\begin{equation}
T_{\mu \nu} = (\epsilon + p)\, u_{\mu} u_{\nu} - p\, g_{\mu\nu},
\label{eq:perfectfluid}
\end{equation}
where $\epsilon$ is the energy density, $p$ the pressure, $u_\mu$ the four-velocity, and $g_{\mu\nu}$ the metric tensor.

By solving Einstein’s equations with the total energy--momentum tensor and the metric in Eq.~\eqref{eq:metricsch}, and imposing $u_\mu = (1,0,0,0)$ for the fluid four-velocity, one obtains the following system:
\begin{flalign}
\frac{dp}{dr} &= -\frac{(\epsilon + p)\,\bigl[m + 4\pi r^3 p\bigr]}{r^2 - 2m\,r},\label{eq:eqh}\\
\frac{dm}{dr} &= 4\pi \epsilon\, r^2, \label{eq:mass}\\
\frac{d\nu}{dr} &= \frac{m + 4\pi r^3 p}{r^2 - 2m\,r}, \label{eq:nueq}
\end{flalign}
where $m$, $p$, and $\epsilon$ are understood to be the \emph{total} mass, pressure, and energy density contributed by ordinary matter, DM, and DE:
\begin{flalign}
m &= m_\mathrm{m} + m_{\mathrm{de}} + m_\chi, \\
p &= p_\mathrm{m} + p_{\mathrm{de}} + p_\chi, \\
\epsilon &= \epsilon_\mathrm{m} + \epsilon_{\mathrm{de}} + \epsilon_\chi.
\end{flalign}
These equations, known as the Tolman--Oppenheimer--Volkoff (TOV) equations, describe the stellar structure under gravitational equilibrium.

\subsection{Closure requirements and boundary conditions}
\label{sec:closure_and_BC}

To solve the TOV equations, it is necessary to specify equations of state (EOSs) that relate the pressure and energy density within the stellar interior. In this work, we provide separate EOSs for each component, introduced in Sec.~\ref{sec:EOS}. For the DE, we employ the standard relation $p_{\mathrm{de}} = \omega\,\epsilon_{\mathrm{de}}$.

Once the EOSs for all components are specified, the TOV system comprises five unknown functions: $m(r)$, $p_{\mathrm{m}}(r)$, $p_{\mathrm{de}}(r)$, $p_{\chi}(r)$, and $\nu(r)$. To obtain a complete solution, two additional conditions must be imposed at each radial coordinate.

In the following subsections, we present two distinct strategies to achieve this closure. The first, referred to as the \emph{coupled fluids model}, assigns fractions of the total energy density to each fluid and prescribes how these fractions either remain constant or vary with radius throughout the stellar interior. The second, called the \emph{interacting fluids model}, makes explicit the notion that energy and momentum can be locally exchanged among baryons, DM, and DE, and introduces interaction terms $Q_{i}$ in the conservation equations to describe this flow.

The TOV equations are integrated outward from the origin $r=0$, where $m(0)=0$ and the central pressures are set to arbitrary values $p_{\mathrm{m}}(0)=p_{\mathrm{m}(c)}$, $p_{\mathrm{de}}(0)=p_{\mathrm{de}(c)}$, and $p_{\chi}(0)=p_{\chi (c)}$. The integration proceeds until the \emph{total} pressure $p(r)$ reaches zero, which defines the stellar surface at $r=R$. The function $m(r)$ evaluated at $R$ then gives the gravitational mass $M(R)$. In addition, the boundary condition for $\nu$ is imposed at the surface: 
\begin{equation}
\nu(R)  =  \tfrac{1}{2} \ln \Bigl(1 - \tfrac{2\,M}{R}\Bigr) .
\label{eq:boundary_nu_at_surface}
\end{equation}

\subsection{Model~I: coupled fluids}
\label{sec:model_1}

In the first approach, we introduce fractional parameters that partition the star's total energy density among ordinary matter, DM, and DE. Specifically, we define a fraction $y_{\mathrm{m}}$ for baryonic matter and a complementary fraction $y_{\mathrm{dark}}$ for the total dark component:
\begin{eqnarray}
y_{\mathrm{m}} &=& \frac{\epsilon_{\mathrm{m}}}{\epsilon_{\mathrm{m}} + \epsilon_{\mathrm{dark}}}, 
\label{eq:fraction-y}\\
y_{\mathrm{dark}} &=& \frac{\epsilon_{\mathrm{dark}}}{\epsilon_{\mathrm{m}} + \epsilon_{\mathrm{dark}}}, 
\label{eq:fraction-y_dark}
\end{eqnarray}
where
\begin{equation}
\epsilon_{\mathrm{dark}} = \epsilon_{\mathrm{de}} + \epsilon_{\chi},
\label{eq:definition_e_dark}
\end{equation}
and
\begin{equation}
y_{\mathrm{m}} + y_{\mathrm{dark}} = 1.
\label{eq:closure_y}
\end{equation}
Once either $y_{\mathrm{m}}$ or $y_{\mathrm{dark}}$ is specified, it remains necessary to determine how much of $\epsilon_{\mathrm{dark}}$ is attributed to DM versus DE. To that end, we define the fractions $x_\chi$ and $x_{\mathrm{de}}$ within the dark sector:
\begin{eqnarray}
x_{\chi} &=& \frac{\epsilon_{\chi}}{\epsilon_{\mathrm{dark}}}, \\
x_{\mathrm{de}} &=& \frac{\epsilon_{\mathrm{de}}}{\epsilon_{\mathrm{dark}}}, 
\end{eqnarray}
subject to
\begin{eqnarray}
x_{\chi} + x_{\mathrm{de}} = 1.
\label{eq:closure_x}
\end{eqnarray}
By specifying one of the $y$ fractions and one of the $x$ fractions at each radial coordinate, the stellar structure equations become fully determined. In this work, we explore both fixed and radially varying choices of $x_\chi$ and $y_{\mathrm{m}}$ to investigate how different partitions of DM and DE affect the structure of the neutron star.

In the following, our aim is to rewrite Eq.~\eqref{eq:eqh}, the TOV pressure equation, so that it depends exclusively on the gradient of the baryonic-matter pressure $p_{\mathrm{m}}$. To achieve this, we first isolate the contribution from the dark sector by noting that
\begin{equation}
p_{\mathrm{dark}} = p_{\mathrm{de}} + p_{\chi},
\end{equation}
and by using Eq.~\eqref{eq:definition_e_dark} to express $\epsilon_{\mathrm{dark}}$ in terms of the matter density $\epsilon_{\mathrm{m}}$. In particular, differentiating $p_{\mathrm{dark}}$ yields
\begin{equation}
\begin{split}
\frac{dp_{\mathrm{dark}}}{dr} =& \frac{dp_{\mathrm{de}}}{dr} + \frac{dp_{\chi}}{dr}
= \frac{dp_{\mathrm{de}}}{d\epsilon_{\mathrm{de}}} \,\frac{d\epsilon_{\mathrm{de}}}{dr}
+\frac{dp_{\chi}}{d\epsilon_{\chi}}\,\frac{d\epsilon_{\chi}}{dr}\\[4pt]
=& \frac{dp_{\mathrm{de}}}{d\epsilon_{\mathrm{de}}}\,\frac{d\epsilon_{\mathrm{dark}}}{dr}
+\frac{d\epsilon_{\chi}}{dr}\,\Bigl(\,\frac{dp_{\chi}}{d\epsilon_{\chi}}
-\frac{dp_{\mathrm{de}}}{d\epsilon_{\mathrm{de}}}\Bigr).
\end{split}
\end{equation}
Since $\epsilon_{\mathrm{dark}}$ is related to $\epsilon_{\mathrm{m}}$ via
\begin{equation}
\epsilon_{\mathrm{dark}}
=\epsilon_{\mathrm{m}}\,\frac{(1-y_{\mathrm{m}})}{y_{\mathrm{m}}},
\end{equation}
we can substitute and assume $x_{\chi}$ is constant to derive an expression for the total pressure gradient $dp/dr$ purely in terms of $dp_{\mathrm{m}}/dr$. Straightforward manipulations then give
\begin{equation}
\begin{split}
\frac{dp}{dr}
=\frac{dp_{\mathrm{m}}}{dr}\,
\biggl[\,
1
+\frac{d\epsilon_{\mathrm{m}}}{dp_{\mathrm{m}}}\,
\Bigl(\,\frac{1-y_{\mathrm{m}}}{y_{\mathrm{m}}}
-\frac{\epsilon_{\mathrm{m}}}{y_{\mathrm{m}}^2}\,\frac{dy_{\mathrm{m}}}{d\epsilon_{\mathrm{m}}}
\Bigr)\\
\qquad\qquad\times
\Bigl(\,
\frac{dp_{\mathrm{de}}}{d\epsilon_{\mathrm{de}}}\,\bigl(1 - x_{\chi}\bigr)
+
\frac{dp_{\chi}}{d\epsilon_{\chi}}x_{\chi}
\Bigr)\biggr].
\end{split}
\end{equation}
Consequently, the TOV pressure equation \eqref{eq:eqh} is recast in terms of $\tfrac{dp_{\mathrm{m}}}{dr}$:
\begin{equation}
\begin{aligned}
\frac{dp_{\mathrm{m}}}{dr}
=&-\frac{\bigl[p+\epsilon\bigr]\bigl[m + 4\pi\,r^3\,p\bigr]}{r^2 - 2\,m\,r}\\
&\times
\biggl[\,
1
+\frac{d\epsilon_{\mathrm{m}}}{dp_{\mathrm{m}}}\,
\Bigl(\,\frac{1-y_{\mathrm{m}}}{y_{\mathrm{m}}}
-\frac{\epsilon_{\mathrm{m}}}{y_{\mathrm{m}}^2}\,\frac{dy_{\mathrm{m}}}{d\epsilon_{\mathrm{m}}}
\Bigr)\\
&\quad\times
\Bigl(
\frac{dp_{\mathrm{de}}}{d\epsilon_{\mathrm{de}}}\,\bigl(1 - x_{\chi}\bigr)
+
\frac{dp_{\chi}}{d\epsilon_{\chi}}x_{\chi}
\Bigr)
\biggr]^{-1}.
\label{eq:modified-TOV}
\end{aligned}
\end{equation}

These expressions reduce to simpler forms in the limit of constant $y_{\mathrm{m}}$. Moreover, if $x_{\chi}=0$ ($x_{\mathrm{de}}=1$), only DE contributes along with baryonic matter, while $x_{\chi}=1$ isolates DM as the only dark component. Finally, $y_{\mathrm{m}}=1$ recovers the conventional TOV equation for a purely baryonic star, irrespective of the value of $x_{\chi}$.

\emph{Prescriptions for $y_{\mathrm{m}}$ and $x_{\chi}$.}
As proposed in Ref.~\cite{Araujo:2024txe}, we adopt the following prescription to describe how the baryonic fraction $y_{\mathrm{m}}$ varies with the baryonic energy density $\epsilon_{\mathrm{m}}$:
\begin{equation}
y_{\mathrm{m}}(\epsilon_{\mathrm{m}}) = 1 + \frac{\beta - 1}{1 + \left(\frac{\epsilon_{\star}}{\epsilon_{\mathrm{m}}}\right)^4},
\label{eq:variable_y}
\end{equation}
where $\beta$ and $\epsilon_{\star}$ are free parameters. This functional form ensures that when $\epsilon_{\mathrm{m}} \ll \epsilon_{\star}$, the fraction $y_{\mathrm{m}}(\epsilon_{\mathrm{m}})$ remains close to 1 (dominated by baryons), whereas for $\epsilon_{\mathrm{m}} \gg \epsilon_{\star}$, it smoothly approaches $\beta$. Additionally, if $\epsilon_{\star} = 0$, then $y_{\mathrm{m}}(\epsilon_{\mathrm{m}}) = \beta$ at all densities, resulting in a constant baryonic fraction throughout the star. The parameter $\epsilon_{\star}$ therefore governs the density scale of this transition, while $\beta$ sets the asymptotic value of $y_{\mathrm{m}}$. Notably, if $\beta = 1$, the fraction remains fixed at $y_{\mathrm{m}}(\epsilon_{\mathrm{m}}) = 1$ for all densities, recovering the case of a purely baryonic star (i.e.\ no effective dark sector).

Throughout this work, the dark-sector fraction $x_{\chi} = \epsilon_{\chi}/\epsilon_{\mathrm{dark}}$ is taken to be a fixed constant, independent of radius or energy density. Hence, in any given configuration, one only needs to specify the parameters $\beta$ and $\epsilon_{\star}$ in Eq.~\eqref{eq:variable_y}, along with the chosen value of $x_{\chi}$. In the sections that follow, we explore both constant and variable prescriptions for $y_{\mathrm{m}}$, but always hold $x_{\chi}$ fixed to focus on how the relative fraction of DM versus DE influences the stellar structure.

\subsection{Model~II: interacting fluids with DM--mediated exchange}
\label{sec:model_2}

Unlike the coupled fluids model, which imposes fractional constraints on the energy densities, the interacting fluids approach explicitly allows for energy--momentum transfer among baryonic matter, DM, and DE. To implement this, we begin by enforcing conservation of the \emph{total} energy--momentum tensor,
\begin{align}
T^{\mu \nu}_{\ ; \mu} \equiv 
\frac{\partial T^{\mu \nu}}{\partial x^\mu} 
+\Gamma^\mu_{\mu \lambda}\,T^{\lambda \nu} 
+\Gamma^\nu_{\mu \lambda}\,T^{\mu \lambda}
=0,
\end{align}
where $T^{\mu \nu}$ is the sum of contributions from each fluid:
\begin{align}
T^{\mu \nu} 
= T^{\mu \nu (\mathrm{m})}
    + T^{\mu \nu (\chi)}
    + T^{\mu \nu (\mathrm{de})}.
\end{align}
To incorporate local interactions among these fluids, we introduce exchange terms $Q_\mathrm{m}$, $Q_\chi$, and $Q_\mathrm{de}$ such that
\begin{align}
T^{\mu \nu (\mathrm{m})}_{\ ; \mu} &= Q_\mathrm{m}, 
\label{Tm}\\
T^{\mu \nu (\chi)}_{\ ; \mu}       &= Q_\chi,       
\label{Tx}\\
T^{\mu \nu (\mathrm{de})}_{\ ; \mu}&= Q_\mathrm{de}, 
\label{Tde}
\end{align}
subject to the overall constraint
\begin{align}
Q_\chi = -\bigl(Q_\mathrm{m} + Q_\mathrm{de}\bigr),
\label{eq:equilibrium_of_Q}
\end{align}
which follows from the total conservation law $T^{\mu \nu}_{\ ; \mu}=0$.

Employing the metric in Eq.~\eqref{eq:metricsch} and the perfect-fluid form of each energy--momentum tensor [Eq.~\eqref{eq:perfectfluid}], one obtains the following structure equations for the pressure of each component ($i = \mathrm{m}, \mathrm{de}, \chi$):
\begin{flalign}
\frac{dp_{\mathrm{m}}}{dr}  + \frac{(p_{\mathrm{m}} + \epsilon_{\mathrm{m}}) (m + 4\pi r^3 p)}{r^2 - 2 m r} 
&=  Q_{\mathrm{m}},  \label{eq:interracting_baryons_model2}\\  
\frac{dp_{\mathrm{de}}}{dr}
&=  Q_{\mathrm{de}}, \label{eq:interracting_de_1}  \\
\frac{dp_{\chi}}{dr} 
 + \frac{\bigl(p_{\chi} + \epsilon_{\chi}\bigr) \bigl(m + 4\pi r^3 p\bigr)}{r^2 - 2 m r} 
&= -\bigl(Q_{\mathrm{de}} + Q_{\mathrm{m}}\bigr),  \label{eq:interracting_chi_model2} \\
\frac{dm}{dr} &= 4\pi \epsilon  r^2, \label{eq:mass}\\
\frac{d\nu}{dr} &= \frac{m + 4\pi r^3 p}{r^2 - 2m r}, \label{eq:nueq}
\end{flalign}
where $p$ is the total pressure. In Eq.~\eqref{eq:interracting_de_1}, DE is treated as vacuum energy with $p_{\mathrm{de}} = - \epsilon_{\mathrm{de}}$, which significantly simplifies that particular equation.

Once the EOSs for all components and the exchange terms $Q_{i}$ are specified, the TOV system comprises five equations with five unknown functions:  $p_{\mathrm{m}}(r)$, $p_{\mathrm{de}}(r)$, $p_{\chi}(r)$, $m(r)$, and $\nu(r)$. As before, we integrate outward from $r=0$, where $m(0)=0$ and $p_{\mathrm{m}}(0)=p_{\mathrm{m}(c)}$, $p_{\mathrm{de}}(0)=p_{\mathrm{de}(c)}$, $p_{\chi}(0)=p_{\chi(c)}$. The stellar radius $R$ is identified by the point where the \emph{total} pressure $p(r)$ vanishes, and $m(R)$ then gives the gravitational mass $M(R)$. The boundary condition for $\nu$ is imposed at $r=R$ via Eq. \eqref{eq:boundary_nu_at_surface}.

\textit{Remark on fractional relations.}  Following the procedure used in the coupled fluids model, one may still introduce fractions $y_{\mathrm{m}(c)}$ and $x_{\chi(c)}$ at the star’s center to fix the central pressures $p_{\mathrm{m}(c)}$, $p_{\mathrm{de}(c)}$ and  $p_{\chi(c)}$. However, because $Q_{i}\neq 0$ in this interacting scenario, these fractions do not remain fixed across all the star.  Since the interaction terms $Q_{\mathrm{m}}$, $Q_{\mathrm{de}}$, and $Q_{\chi}$ redistribute energy and momentum, the fractions evolve with the radial coordinate.

To complete this framework, we must specify explicit functional forms for the interaction terms $Q_{i}$. Here, we assume that DM serves as the primary mediator of energy exchange with both baryons and DE. Following cosmological arguments (e.g., Ref.~\cite{caldera2009dynamics}), we posit that whenever $\epsilon_{\chi}$ changes with radius, a fraction of that change is transferred to the other fluids. Concretely, we define
\begin{equation}
Q_i = \alpha \frac{\epsilon_i}{\epsilon_\chi} \frac{d\epsilon_\chi}{dr},
\label{eq:interaction}
\end{equation}
where $i \in \{\mathrm{m}, \mathrm{de}\}$ and $\alpha$ is a dimensionless coupling constant.

By construction, Eq.~\eqref{eq:interaction} ensures that energy is exchanged in proportion to the relative energy densities of the non--DM fluids. In conjunction with the global conservation condition in Eq.~\eqref{eq:equilibrium_of_Q}, the net flux from DM becomes
\begin{equation}
Q_\chi 
= -\,\bigl(Q_{\mathrm{m}} + Q_{\mathrm{de}}\bigr)
=  -\,\alpha\,\frac{\epsilon_{\mathrm{m}} + \epsilon_{\mathrm{de}}}{\epsilon_\chi}\,
\frac{d\epsilon_\chi}{dr}.
\label{eq:interaction_2222}
\end{equation}

\subsection{Model~III: interacting fluids with a unified dark sector}
\label{sec:model_3}

In this alternative formulation, DM and DE are combined into a single dark sector that exchanges energy only with baryonic matter. Unlike Model~II, which introduces separate interaction terms $Q_{\mathrm{m}}$, $Q_{\mathrm{de}}$, and $Q_{\chi}$, we now employ a single exchange parameter $Q$ linking baryons and the unified dark sector:
\begin{align}
T^{\mu \nu (\mathrm{m})}_{\ ;\mu} &= Q, \\
T^{\mu \nu (\mathrm{dark})}_{\ ;\mu} &= - Q,
\end{align}
where $T^{\mu \nu (\mathrm{dark})} = T^{\mu \nu (\chi)} + T^{\mu \nu (\mathrm{de})}$ encompasses both DM and DE.

Following the same procedure as in Model~II, one obtains the TOV equations for $p_{\mathrm{m}}$ and $p_{\mathrm{dark}}$:
\begin{flalign}
\frac{dp_{\mathrm{m}}}{dr} 
 + \frac{\bigl(p_{\mathrm{m}} + \epsilon_{\mathrm{m}}\bigr)\bigl(m + 4\pi r^3 p\bigr)}{r^2 - 2 m r} 
&=  Q, 
\label{eq:dp_m_unified}\\
\frac{d p_{\mathrm{dark}}}{dr} 
 + \frac{\bigl(p_{\chi} + \epsilon_{\chi}\bigr)\bigl(m + 4\pi r^3 p\bigr)}{r^2 - 2 m r}
&= - Q, 
\label{eq:dp_dark_unified}\\
\frac{dm}{dr} &= 4\pi \epsilon r^2, 
\label{eq:mass_unified}\\
\frac{d\nu}{dr} &= \frac{m + 4\pi r^3 p}{r^2 - 2 m r}, 
\label{eq:nueq_unified}
\end{flalign}
where $p = p_{\mathrm{m}} + p_{\mathrm{dark}}$ is the total pressure, $\epsilon = \epsilon_{\mathrm{m}} + \epsilon_{\mathrm{dark}}$ is the total energy density, and Eq.~\eqref{eq:dp_dark_unified} is simplified using $p_{\mathrm{de}} = - \epsilon_{\mathrm{de}}$.

To specify how energy flows between the dark sector and baryons, we assume
\begin{equation}
Q =  \alpha \frac{d\epsilon_{\mathrm{dark}}}{dr} = \alpha \frac{d\epsilon_{\mathrm{dark}}}{d p_{\mathrm{dark}}} \frac{d p_{\mathrm{dark}}}{dr},
\label{eq:Q_definition_model3}
\end{equation}
where $\alpha$ is a dimensionless coupling constant. Changes in $\epsilon_{\mathrm{dark}}$ at a given radius thus drive net energy flow $Q$ to or from the baryons.

Although DM and DE are merged into a single fluid in terms of $Q$, we still distinguish their partial pressures $p_{\chi}$ and $p_{\mathrm{de}}$ (and corresponding densities) to evaluate the EOS. As in the coupled fluids model (Model~I), we specify $x_{\chi} \equiv \epsilon_{\chi}/\epsilon_{\mathrm{dark}}$ to partition the dark sector between matterlike and vacuumlike components. This extra condition serves the same purpose as the fractional relations in Model~I, ensuring that the system is well-defined even though there is only one interaction term $Q$. In this sense, Model~III is a hybrid of the coupled and interacting approaches: the dark sector interacts with baryons as a unified fluid, yet we retain an internal partition via $x_{\chi}$.

As with the other models, we integrate Eqs.~\eqref{eq:dp_m_unified}--\eqref{eq:nueq_unified} from $r=0$ outward, imposing $m(0)=0$ and initial pressures $p_{\mathrm{m}}(0)=p_{\mathrm{m}(c)}$, $p_{\mathrm{dark}}(0)=p_{\mathrm{dark}(c)}$. The radius $R$ is defined by $p(R)=0$, and $m(R)$ then gives the gravitational mass $M(R)$. The boundary condition for $\nu$ is imposed at $r=R$ via Eq.~\eqref{eq:boundary_nu_at_surface}. Throughout this procedure, $x_{\chi}$ remains a fixed parameter at every radius.

\begin{figure*}[tbh]
\centering
\includegraphics[width=\textwidth]{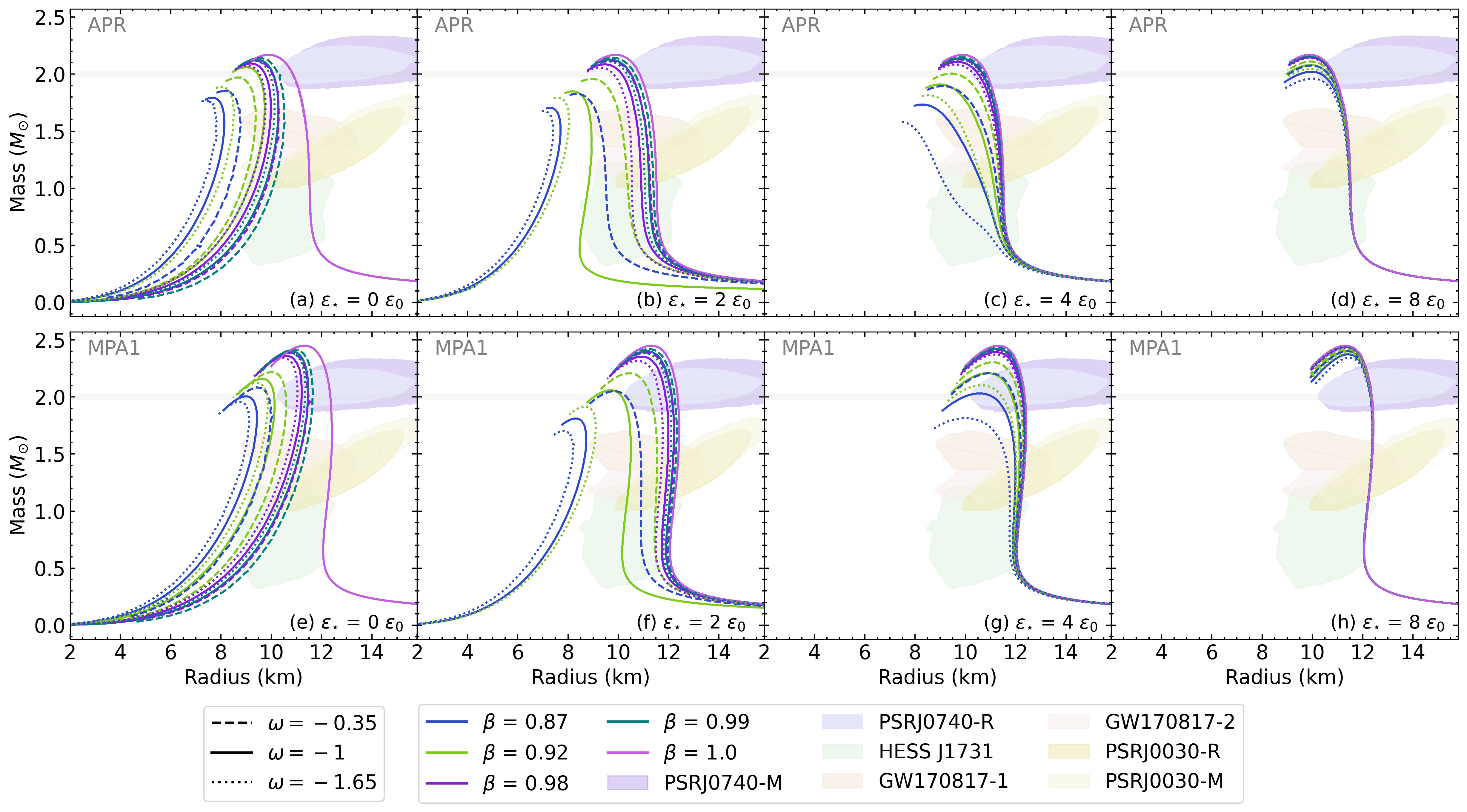}
\caption{
Mass--radius curves for \emph{Model~I} in the DE-only limit (no DM), as described in Sec.~\ref{sec:darkEnergyOnly}. 
Results are shown for two representative baryonic EOSs: APR (top panels) and MPA1 (bottom panels). 
In each row, panels (a)--(d) (and (e)--(h) for MPA1) correspond to increasing threshold densities $\epsilon_{\star}$ at which DE becomes significant. 
Three line types represent different DE EOSs: $\omega=-1.65$ (phantom-like, dotted), $\omega=-1$ (vacuum, dashed), and $\omega=-0.35$ (dash-dotted). 
Different colors indicate various values of the fraction $\beta$. 
Shaded bands show observational constraints from pulsars and gravitational-wave data.
}
\label{fig:1}
\end{figure*}

\begin{figure*}[tbh]
\centering
\includegraphics[width=\textwidth]{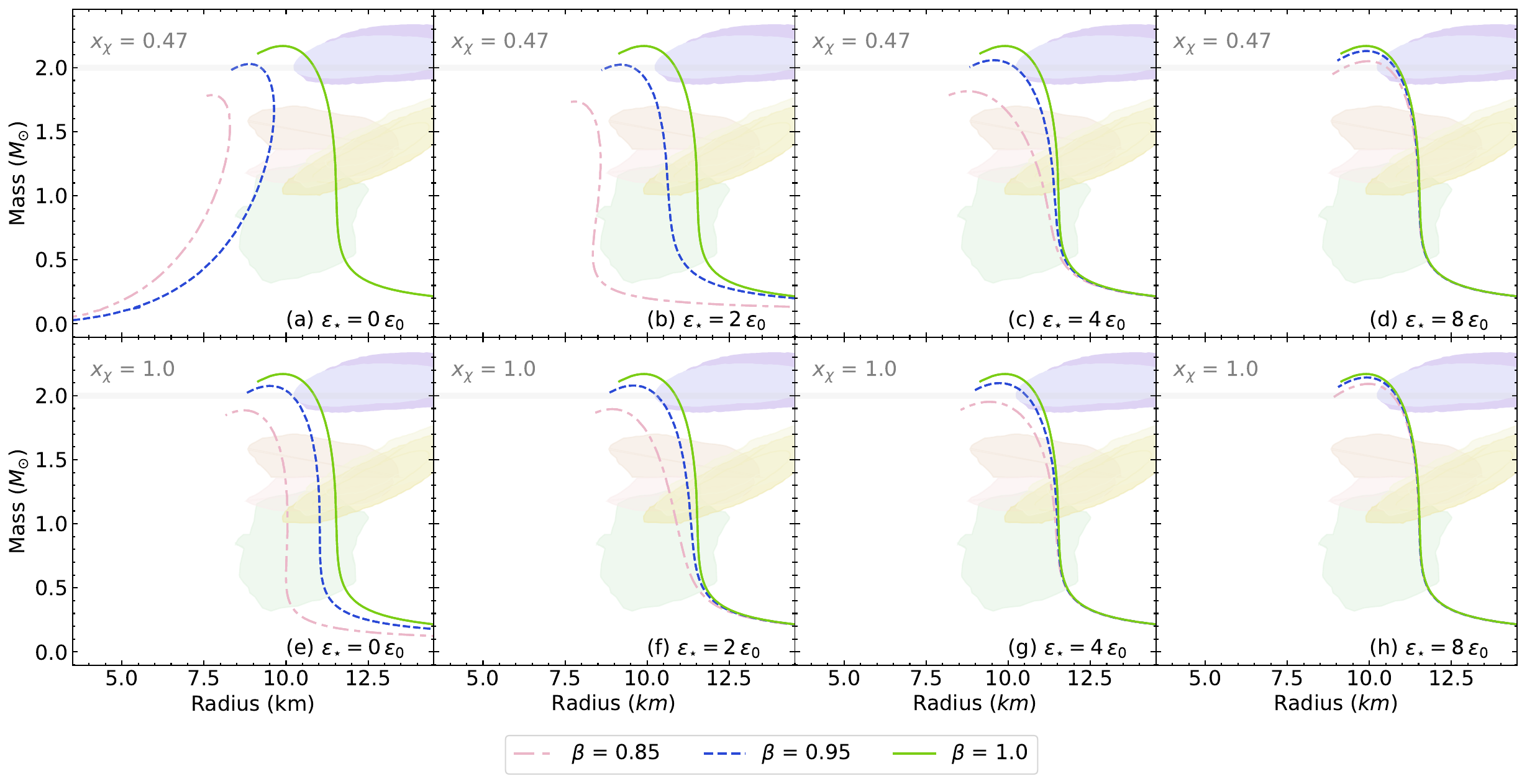}
\caption{Mass--radius relations for Model~I in the presence of both dark energy (DE) and dark matter (DM), as discussed in Sec.~\ref{sec:coupledNoInteraction}. We employ the APR EOS, treating DM as a degenerate Fermi gas, and assume a dark-matter particle mass of $m_\chi=10\,\mathrm{GeV}$ with no self-interaction ($q=0$, $\alpha=0$). Each panel corresponds to fixed values of the baryonic fraction $y_{\mathrm{m}}$ and the DM fraction $x_{\chi}$, while the curves also span threshold densities $\epsilon_{\star}=0,\,2,\,4,\,8\,\epsilon_0$, which control the onset of dark energy. The shaded bands have the same meaning as in Fig.~\ref{fig:1}, representing observational constraints from pulsars and gravitational-wave measurements.}
\label{fig:2}
\end{figure*}

\section{Results}
\label{sec:results}
In this section, we investigate how DE and DM modify neutron star structure under the three main scenarios presented in Sec.~\ref{sec:structure} (Models I, II, and III). We use the EOSs described in Sec.~\ref{sec:EOS} for each fluid component and assess the resulting stellar configurations, such as mass–radius relations, against current observational constraints, including high-precision pulsar mass measurements and gravitational-wave data from binary neutron-star mergers.

\subsection{Results of Model~I: coupled fluids}
\label{sec:coupled_fluids}

\subsubsection{Dark energy only}
\label{sec:darkEnergyOnly}

In this subsection, we begin by examining a special case of Model~I in which there is no DM ($x_{\chi}=0$). This scenario provides a direct extension of our earlier work (Ref.~\cite{Araujo:2024txe}) that focused on the role of DE in neutron-star interiors. Our goal here is to study how varying the DE EOS, specifically the parameter $\omega$, influences stellar structure.

Figure~\ref{fig:1} extends the results of Ref.~\cite{Araujo:2024txe} by showing how changes in $\omega$ shift the mass--radius curves. Three main trends, already noted in Ref.~\cite{Araujo:2024txe}, remain evident:

\begin{itemize}

\item Increasing $\epsilon_{\star}$ confines DE to higher densities, causing the star to behave more like a standard neutron star with a softer EOS that yields lower maximum masses and smaller radii. For $\epsilon_{\star} = 0$ or sufficiently small $\beta$, DE becomes relevant at very low pressures, and the resulting mass--radius curves resemble those of strange quark stars.

\item Decreasing the baryonic fraction $y_{\mathrm{m}}$ (i.e.\ smaller $\beta$) systematically reduces both the maximum mass and the corresponding radius.

\item Moderate $\beta$ often aligns well with current data, while large DE fractions tend to move the theoretical curves away from observed mass--radius measurements.
\end{itemize}

What differs here from our previous study is the explicit analysis of $\omega$ itself. At fixed $\beta$, lowering $\omega$ below $-1$ (a phantom-like EOS) yields slightly larger masses at a given radius, although these remain below those of a purely baryonic configuration ($\beta=1$). In addition, very negative values of $\omega$ may destabilize the star if the dark sector dominates. These effects appear qualitatively similar for both the APR and MPA1 EOSs.

\subsubsection{Dark energy and dark matter}
\label{sec:coupledNoInteraction}

We next incorporate DM, modeled as a degenerate Fermi gas, and allow both DM and DE to coexist without explicit interactions. Instead, we use the coupling relations from Eq.~\eqref{eq:modified-TOV}, which specify how the total energy density is partitioned at each radius between baryons ($y_{\mathrm{m}}$) and the dark sector ($x_{\chi}$ for DM vs.\ DE).

Figure~\ref{fig:2} displays the resulting mass--radius curves for $m_{\chi} = 10\,\mathrm{GeV}$, assuming the APR EOS for ordinary matter. Each column corresponds to a threshold density $\epsilon_{\star}=0,\,2,\,4,\,8\,\epsilon_0$, indicating at which densities the dark components become relevant. The top row [Figs.~\ref{fig:2}(a)--(d)] adopts $x_{\chi}=0.47$, meaning the dark sector is composed of approximately half DM and half DE, while the bottom row [Figs.~\ref{fig:2}(e)--(h)] sets $x_{\chi}=1.0$, indicating a purely DM composition. In both rows, each panel compares three baryonic fractions $y_{\mathrm{m}} = 0.85,\,0.95,\,1.0$.

\paragraph{Baryonic fraction $y_{\mathrm{m}}$.}
Lowering $y_{\mathrm{m}}$ from 1.0 to 0.85 increases the fraction of dark components and, as a result, reduces both the maximum mass and radius. With fewer baryons providing pressure support, gravity is less effectively balanced—especially if part of the dark sector exerts negligible or negative pressure. Consequently, the curves for $y_{\mathrm{m}}=0.85$ shift toward lower masses at a fixed radius (or smaller radii at a fixed mass), compared to $y_{\mathrm{m}}=1.0$.

\paragraph{Dark-sector composition $x_{\chi}$.} 
Contrasting the top row ($x_{\chi}=0.47$) with the bottom row ($x_{\chi}=1.0$) reveals a markedly stronger influence of the dark sector when a significant fraction of dark energy is present. In the $x_{\chi}=0.47$ case, where dark energy contributes substantially, the mass–radius curves are shifted to lower masses and radii, to the extent that for $\epsilon_{\star}=0$ the solutions may even become self-bound, exhibiting a mass–radius profile reminiscent of strange quark stars. This behavior results from the negative pressure associated with dark energy, which causes the star's surface (i.e., the radius at which the total pressure vanishes) to occur at a finite energy density. In contrast, when $x_{\chi}=1.0$,  the relatively heavy dark matter particles soften the EOS, but the absence of a negative pressure component means that the overall softening of the EOS is much less pronounced; in this regime, the mass–radius curves do not exhibit self-bound behavior.

\paragraph{Threshold density $\epsilon_{\star}$.} 
Each column in Fig.~\ref{fig:2} corresponds to a different threshold density, namely $\epsilon_{\star}=0,\,2,\,4,\,8\,\epsilon_0$. Increasing $\epsilon_{\star}$ delays the onset of the dark components to higher densities, so that the star behaves more like a conventional neutron star in its low-density regions. Consequently, for larger $\epsilon_{\star}$ the M--R curves approach those of a purely baryonic star, typically exhibiting higher maximum masses and slightly larger radii. The most striking behavior occurs when $\epsilon_{\star}=0$. In the mixed scenario ($x_{\chi}=0.47$), the presence of dark energy leads to self-bound configurations. In contrast, when $x_{\chi}=1.0$ and $\epsilon_{\star}=0$, the softening of the EOS is maximal and produces the largest shift in the M--R curves, albeit without yielding self-bound objects.

\subsection{Results of Model~II: interacting fluids with DM--mediated exchange}
\label{sec:interactingFluids}

\begin{figure}[tb]
\centering
\includegraphics[width=0.48 \textwidth]{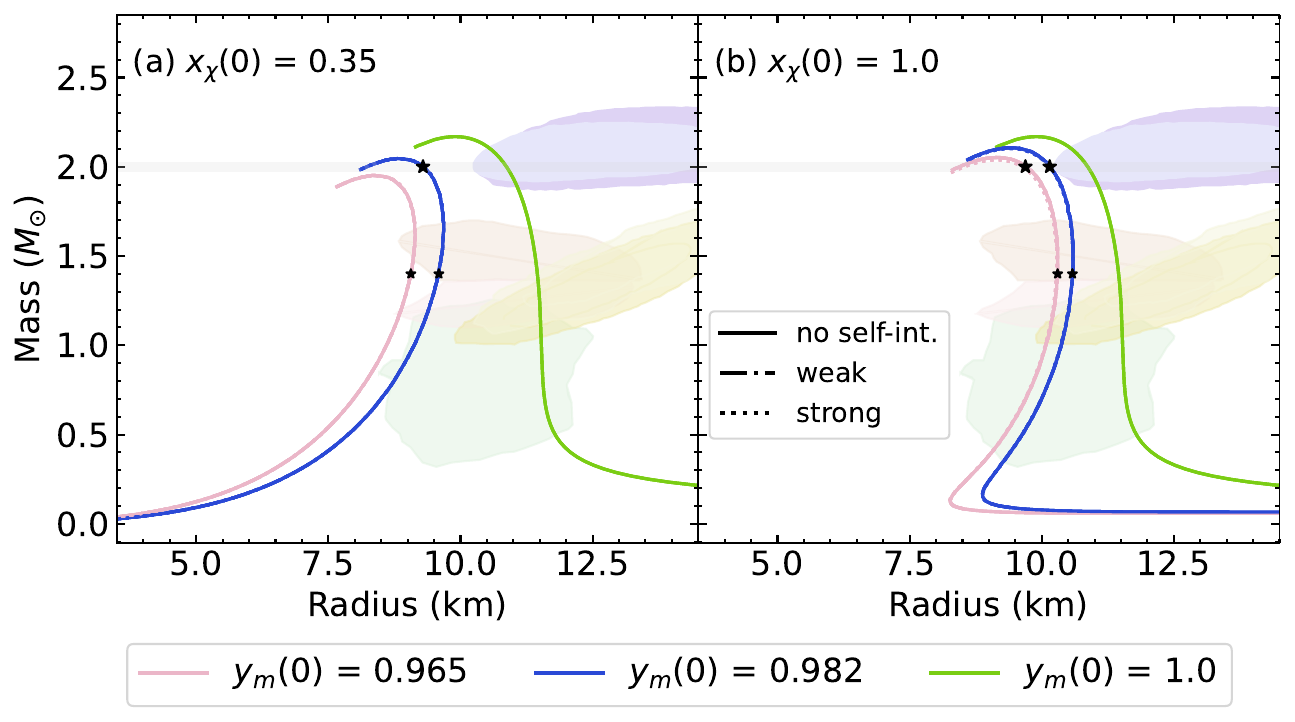}
\caption{Mass--radius relationships for Model~II under the APR EOS, illustrating the interplay of dark energy (DE), dark matter (DM), and DM self-interactions. Panel (a) shows the case with $x_{\chi}(0)=0.35$, whereas panel (b) corresponds to $x_{\chi}(0)=1$, representing a purely DM configuration at $r=0$. Three central baryonic fractions $y_{\mathrm{m}}(0)$ are depicted in different colors, and each is examined under no self-interaction, weak self-interaction ($q=0.1$), and strong self-interaction ($q=1000$). Note that this is the only figure in the article where DM self-interactions are considered. We assume a dark-matter particle mass of $m_\chi=10\,\mathrm{GeV}$ and set $\alpha=1$ in Eq.~\eqref{eq:interaction}. The asterisks mark the $1.4\,M_{\odot}$ and  $2.0\,M_{\odot}$ configurations, which are analyzed in the subsequent figures.}
\label{fig:3}
\end{figure}

\begin{figure}
\centering    
\includegraphics[width=0.48\textwidth]{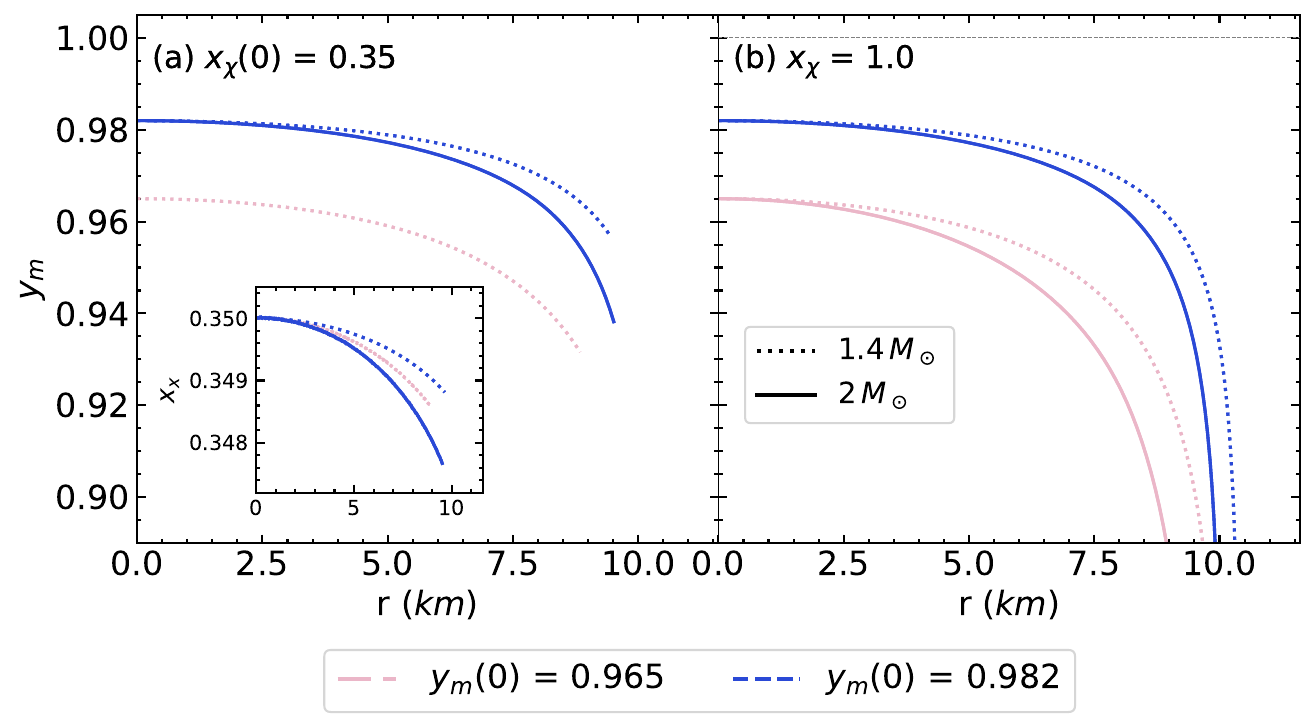}
\caption{Radial profiles of the baryonic fraction $y_{\mathrm{m}}(r)$ and the DM fraction $x_{\chi}(r)$ in neutron stars of $1.4\,M_{\odot}$ (canonical) and $2.0\,M_{\odot}$ (high mass) for Model~II. }
\label{fig:4}
\end{figure}

\begin{figure}[tb]
\centering    
\includegraphics[width=0.48\textwidth]{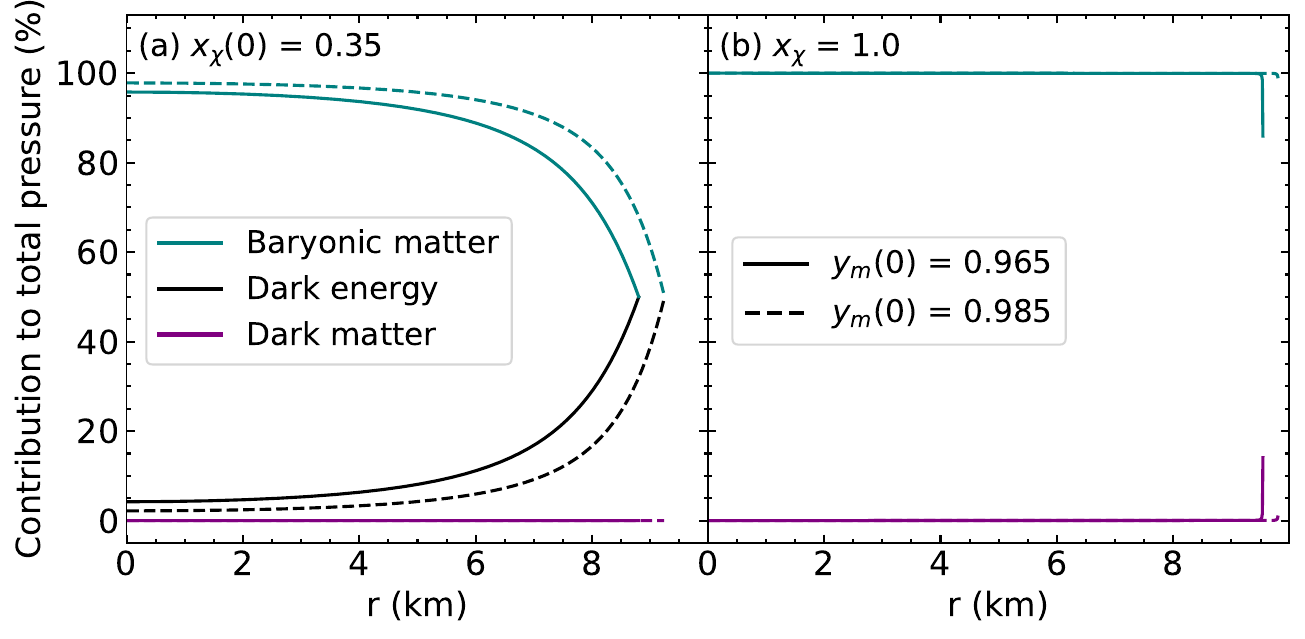}
\caption{Pressure profiles (as a percentage of the total pressure) versus radial coordinate for neutron stars in Model~II for the same cases discussed in Fig.~\ref{fig:4}. }
\label{fig:5}
\end{figure}

\begin{figure}[tb]
\centering    
\includegraphics[width=0.48 \textwidth]{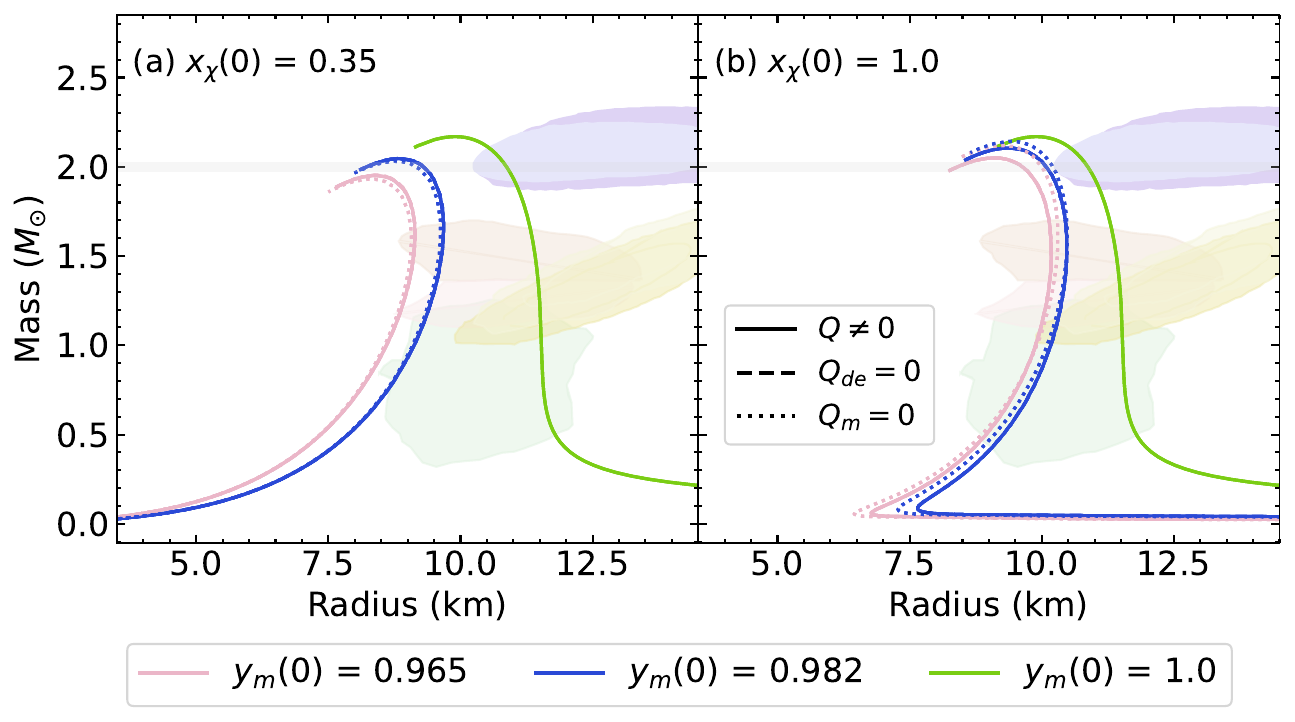}
\caption{Mass--radius relations for Model~II, as described in Sec.~\ref{sec:model_2}, under various scenarios for the interaction terms. Dark matter particles are assumed to have a mass of $m_\chi=10\,\mathrm{GeV}$ with no self-interaction, and we set $\alpha=1$ in the interaction term of Eq.~\eqref{eq:interaction}. See text for further discussion.}
\label{fig:6}
\end{figure}

\begin{figure}[tb]
\centering    
\includegraphics[width= 0.48 \textwidth]{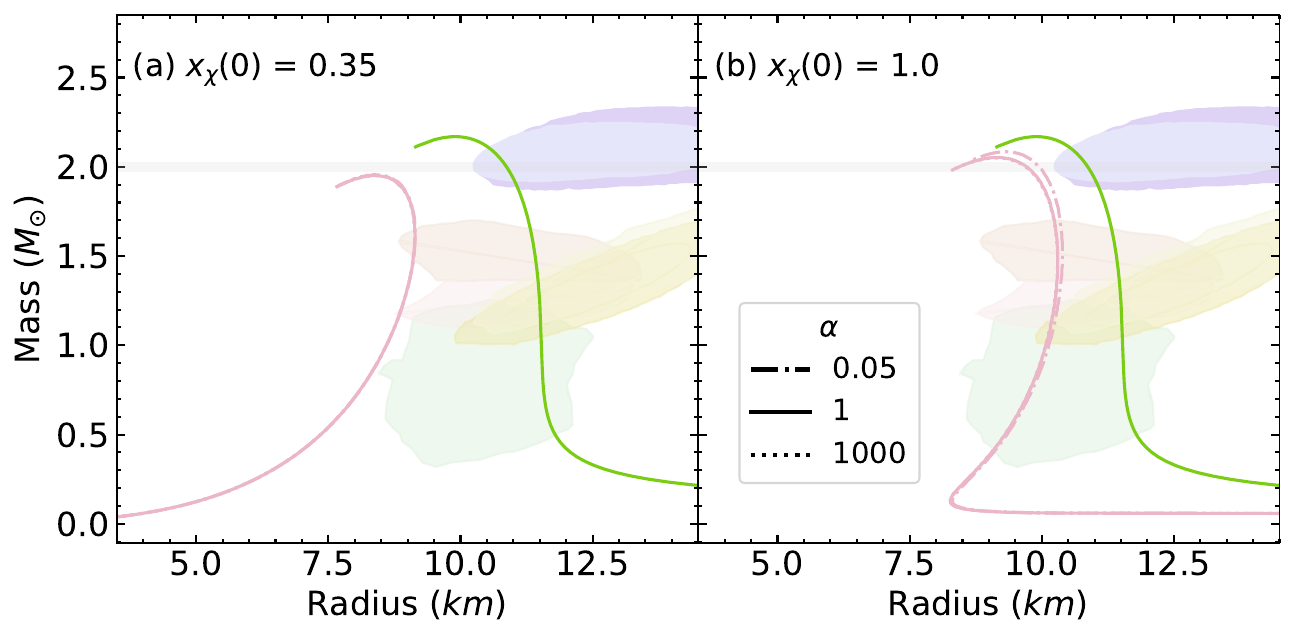}    
\caption{Mass–radius relations predicted by Model~II for two initial dark‑sector fractions: (a) $x_{\chi}(0)=0.35$ and (b) $x_{\chi}(0)=1.0$.  Each panel displays three stellar sequences obtained for the values of the coupling parameter $\alpha$ listed in the legend.  The baryonic fraction $y_{\mathrm{m}}$ is fixed at 0.965 to ensure a predominantly baryonic configuration.}
\label{fig:7}
\end{figure}

In Figs.~\ref{fig:3}--\ref{fig:7}, we explore various aspects of Model~II, focusing on how DM self-interactions and the relative fractions of baryons, DM, and DE shape neutron-star configurations.

Figure~\ref{fig:3} displays mass–radius relations computed with the APR EOS, highlighting the impact of varying dark‑sector composition and dark‑matter self‑interactions. Panels (a) and (b) correspond to central dark‑matter fractions $x_{\chi}(0)=0.35$ (mixed DM–DE) and $x_{\chi}(0)=1.0$ (purely DM), respectively. In each panel, we plot three curves for baryonic fractions $y_{\mathrm{m}}(0)$ distinguished by color. Note that $x_{\chi}(0)$ and $y_{\mathrm{m}}(0)$ specify only the composition at $r=0$, but these fractions evolve radially due to local fluid interactions.  In both scenarios, reducing the baryonic fraction shifts the M–R curves toward smaller radii at fixed mass and lowers the maximum supported mass. Physically, this softening arises because the dark sector—whether massive fermionic dark matter with negligible pressure or negative‑pressure dark energy—contributes to the gravitational pull (via its energy density) but offers far less pressure support than ordinary matter. The resulting decrease in the effective stiffness of the combined EOS thus reduces both the star’s radius and its maximum mass.

Figure~\ref{fig:3} also examines the impact of repulsive DM self-interactions and shows that, even for an extreme choice $q=1000$, the mass–radius curves are scarcely altered.  The reason becomes clear by comparing the self-interaction shift $\Delta\epsilon_\chi=\Delta p_\chi$ in Eq. \eqref{eq:vector_contribution_DM}  with the free–Fermi–gas terms $\epsilon_\chi^{\mathrm{FG}},p_\chi^{\mathrm{FG}}$ in Eq. \eqref{eq:FG_for_z_ll_1}.  In the degenerate, non-relativistic regime relevant for a $10~\mathrm{GeV}$ fermion inside a neutron star ($z=k_{F\chi}/m_\chi \lesssim 10^{-2}$), one finds
\begin{align}
\frac{\Delta\epsilon_\chi}{\epsilon_\chi^{\mathrm{FG}}} &= \frac{q^{2}\,z^{3}}{3\pi^{2}} \lesssim 10^{-7}\,q^{2},\\
\frac{\Delta p_\chi}{p_\chi^{\mathrm{FG}}} &= \frac{5\,q^{2}\,z}{3\pi^{2}} \lesssim 10^{-3}\,q^{2}.
\end{align}
Thus, the contribution of self‑interactions to $\epsilon_\chi$ remains negligible, whereas the pressure can indeed be dominated by self‑interactions.

Although a large $q$ can locally make the self‑interaction term dominate the DM pressure, its contribution to the \emph{total} pressure remains negligible.  With $m_\chi=10\;\mathrm{GeV}$, $z\sim10^{-2}$, and central baryonic pressures $p_{\mathrm m}\sim10^{2}$–$10^{3}\;\mathrm{MeV}/\mathrm{fm}^3$, one finds
\begin{equation}
\frac{p_\chi}{p_{\mathrm m}}  \sim \frac{q^2 m_\chi^{4}\,z^{6}}{9\pi^{4}\,p_{\mathrm m}}
\sim q^2 \times  \left(10^{-8}\text{--}10^{-9} \right).
\end{equation}
Hence, even for $q=1000$, the self‑interaction term makes only a negligible contribution to the total pressure. Consequently, DM self‑interactions leave the global stellar structure essentially unchanged, as confirmed by Fig.~\ref{fig:3}.

Figure~\ref{fig:4} illustrates how the baryonic fraction $y_{\mathrm{m}}(r)$ and the DM fraction $x_{\chi}(r)$ evolve from the center ($r=0$) to the stellar surface ($r=R$) for two representative neutron-star configurations: one with $1.4\,M_{\odot}$ (canonical mass) and one with $2.0\,M_{\odot}$ (high mass). These models are indicated by asterisks in Fig.~\ref{fig:3}.

Figure~\ref{fig:4}(a) shows that as $r$ increases, the baryonic fraction $y_{\mathrm{m}}(r)$ decreases, which means the total dark fraction $y_{\mathrm{dark}}(r) = 1 - y_{\mathrm{m}}(r)$ rises accordingly. Within the dark sector, $x_{\chi}(r)$ also declines outward, implying that the fraction of DE $x_{\mathrm{de}}(r) = 1 - x_{\chi}(r)$ grows toward the surface.
These results can be interpreted in terms of the interaction terms $Q_{i}$. From Eqs.~\eqref{eq:interaction} and \eqref{eq:interaction_2222}, one finds
\begin{equation}
Q_{\mathrm{m}} < 0, 
\quad Q_{\mathrm{de}} < 0, 
\quad \text{and} \quad Q_{\chi} > 0, 
\end{equation}
whenever $d\epsilon_{\chi}/dr < 0$. Because $Q_{\mathrm{m}}<0$, baryonic matter loses pressure and energy more rapidly than it would without interaction, so $y_{\mathrm{m}}(r)$ diminishes with $r$, as shown in Fig.~\ref{fig:4}(a). Correspondingly, $y_{\mathrm{dark}}(r)$ increases. On the other hand, $Q_{\mathrm{de}}<0$ implies that DE’s pressure, already negative at the center, becomes even more negative with increasing $r$ (see Eq.~\eqref{eq:interracting_de_1}). A more negative pressure leads to an increased DE density $\epsilon_{\mathrm{de}}(r)$ near the surface. Hence, $x_{\mathrm{de}}(r) = 1 - x_{\chi}(r)$ rises at large $r$, consistent with the inset in Fig.~\ref{fig:4}(a). Finally, $Q_{\chi}>0$ indicates that DM loses pressure and energy density less rapidly than it would otherwise, so $\epsilon_{\chi}(r)$ decreases more slowly. However, since DE grows at an even faster rate, $x_{\chi}(r)$ can still diminish in the outer layers, since it is a fraction of the total dark component. In other words, although DM gains energy through these interactions, the larger increase in DE ultimately causes $x_{\chi}(r)$ to decline. 

The fact that the star’s outer layers contain a higher fraction of dark energy than its core explains why the M–R curves in Fig.~\ref{fig:3}(a) mimic those of strange quark stars. The negative pressure of dark energy, which grows more pronounced near the surface, causes the radius at which the total pressure vanishes to occur at a smaller value than in purely baryonic or predominantly dark-matter configurations, ultimately causing these objects to become self-bound.

In Fig. \ref{fig:4}(b) we consider the case $x_{\chi}(0)=1$, so that no dark energy is present at the center ($\epsilon_{\mathrm{de}}(0)=0$). From Eq. \eqref{eq:interaction}, the exchange term $Q_{\mathrm{de}}=\alpha\,\frac{\epsilon_{\mathrm{de}}}{\epsilon_{\chi}}\frac{d\epsilon_{\chi}}{dr}$ vanishes wherever $\epsilon_{\mathrm{de}}=0$.  Hence, no dark‑energy density can be generated in any subsequent shell, and $\epsilon_{\mathrm{de}}(r)\equiv 0$ throughout the star.  Accordingly, $x_{\chi}(r)=1$ everywhere up to the stellar surface.
In this case, the radial variation of $y_{\mathrm{m}}(r)$ is governed solely by $Q_{\mathrm{m}}<0$, implying that the baryonic fraction decreases with increasing $r$, thereby transferring part of the energy budget into DM. This absence of a DE component in the outer layers explains why the mass--radius behavior in Fig.~\ref{fig:3}(b) resembles that of hadronic stars rather than self-bound stars.

Figure~\ref{fig:5} shows the radial decomposition of the total pressure into baryonic, dark‑matter, and dark‑energy contributions under Model~II, extending the results of Figs.~\ref{fig:3} and \ref{fig:4}. In Fig.~\ref{fig:5}(a), with $x_{\chi}(0)=0.35$, dark matter accounts for a sizable fraction of the energy density but contributes negligibly to the pressure. This reflects our choice of heavy dark‑matter particles ($m_{\chi}=10~\mathrm{GeV}$), which yield high energy density yet very little pressure. Accordingly, the dark energy component dominates the dark sector pressure profile, and its large (negative) surface pressure explains why the M--R curves in Fig.~\ref{fig:3}(a) acquire a self‑bound appearance. In contrast, Fig.~\ref{fig:5}(b) with $x_{\chi}(0)=1.0$ contains only dark matter and no dark energy at the center. Here again, the massive dark matter particles produce minimal pressure but still soften the overall EOS through their energy density contribution. In the absence of a negative pressure component, the resulting pressure partition yields the more hadronic‑star–like mass–radius behavior seen in Fig.~\ref{fig:3}(b).

Figure~\ref{fig:6} compares M--R curves for Model~II under various energy-exchange constraints. The values of $y_{\mathrm{m}}$ and $x_{\chi}(0)$ are the same as those used in Figs.~\ref{fig:3}, \ref{fig:4}, and \ref{fig:5}. Different line styles represent specific restrictions on the interaction terms: the case with $Q_{\mathrm{m}}=0$ corresponds to no energy transfer from dark matter to baryons,  $Q_{\mathrm{de}}=0$ denotes the absence of energy exchange from dark matter to dark energy,  whereas those with $Q\neq0$ reflect the full Model~II scenario.
In the case $x_{\chi}(0)=0.35$, setting $Q_{\mathrm{m}}=0$ produces a slight reduction in both maximum mass and radius, whereas disabling $Q_{\mathrm{de}}$ has a negligible effect. This insensitivity indicates that, when dark energy comprises a significant fraction of the dark sector, the global stellar structure depends only weakly on the details of energy transfer. By contrast, for $x_{\chi}(0)=1.0$, setting $Q_{m}=0$ induces larger changes in mass and radius, especially when $y_\mathrm{m}$ is smaller. This arises because, without energy exchange, the dark sector is less effective at softening the EOS.

Figure ~\ref{fig:7} explores the sensitivity of Model~II to the coupling constant $\alpha$. When the dark sector is a DM–DE mixture, $x_{\chi}(0)=0.35$ [panel~(a)], the three mass–radius sequences for $\alpha=0.05,\,1,\,$ and $\,1000$ are virtually indistinguishable. Even in the purely–DM case, $x_{\chi}(0)=1.0$ [panel~(b)], increasing $\alpha$ produces only a modest shift toward slightly smaller radii and lower maximum masses.

To clarify this weak dependence, we rewrite the DM structure equation (Eq.~\eqref{eq:interracting_chi_model2}) as  
\begin{equation}
\left[1+\alpha\frac{\epsilon_{\mathrm{m}}+\epsilon_{\mathrm{de}}}{\epsilon_\chi} \frac{d\epsilon_\chi}{dp_\chi} \right] \frac{dp_\chi}{dr}  =-\frac{(p_\chi+\epsilon_\chi)(m+4\pi r^3 p)}{r^2-2mr}\;.
\label{eq:TOV_DM_recast_alpha}
\end{equation}
For non‑relativistic fermions of mass $m_\chi=10\ \mathrm{GeV}$,
\begin{equation}
p_\chi \simeq \frac{k_{F\chi}^5}{15\pi^2m_\chi},\quad  \epsilon_\chi \simeq \frac{m_\chi k_{F\chi}^3}{3\pi^2}\;,
\end{equation}
so that $k_{F\chi}\ll m_\chi$, which is the relevant case in neutron stars, implies 
\begin{equation}
\frac{d\epsilon_\chi}{dp_\chi}\approx\frac{5m_\chi^2}{k_{F\chi}^2}\gg1\;.
\label{eq:alphakappa_ll_1}
\end{equation}
Defining  
\begin{equation}
\Upsilon(r)=1+\alpha\frac{\epsilon_{\mathrm{m}}+\epsilon_{\mathrm{de}}}{\epsilon_\chi} \frac{d\epsilon_\chi}{dp_\chi}\gg1,
\end{equation}
Eq.~\eqref{eq:TOV_DM_recast_alpha} reduces to
\begin{equation}
\frac{dp_\chi}{dr} \simeq-\frac{1}{\Upsilon}\,\frac{(p_\chi+\epsilon_\chi)(m+4\pi r^3 p)}{r^2-2mr}\;,
\end{equation}
so that increasing $\alpha$ both enlarges $\Upsilon$ and suppresses $dp_\chi/dr$ by the same factor.  As a consequence, the product $\alpha \, {dp_\chi}/{dr}$, which governs the net energy flux, remains comparable in magnitude to the hydrostatic term. In other words, the exchange terms $Q_i$ self‑regulate, rendering their overall effect effectively independent of $\alpha$.

In the same way, the dark energy structure equation in Model~II, Eq.~\eqref{eq:interracting_de_1}, can be written (for $\Upsilon\gg1$) as
\begin{align}
\frac{dp_\mathrm{de}}{dr}
&=\alpha\frac{\epsilon_\mathrm{de}}{\epsilon_\chi}\frac{d\epsilon_\chi}{dr} =\alpha\frac{\epsilon_\mathrm{de}}{\epsilon_\chi} \frac{d\epsilon_\chi}{dp_\chi}\frac{dp_\chi}{dr} \\
&= -\alpha\frac{\epsilon_\mathrm{de}}{\epsilon_\chi}\, \frac{d\epsilon_\chi}{dp_\chi}\, \frac{1}{\Upsilon}\, \frac{(p_\chi+\epsilon_\chi)(m+4\pi r^{3}p)}  {r^{2}-2mr} \\
&\simeq -\,\frac{\epsilon_{\mathrm{de}}}{\epsilon_{m}+\epsilon_{\mathrm{de}}} \;\frac{(p_\chi+\epsilon_\chi)\,(m+4\pi r^3p)}{r^2-2mr},
\end{align}
which makes manifest that the DE pressure gradient is independent of $\alpha$.

Similarly, applying the same $\Upsilon\gg1$ approximation to the baryonic TOV equation in Model~II [Eq.~\eqref{eq:interracting_baryons_model2}] yields
\begin{align}
\frac{dp_{\mathrm{m}}}{dr}
&\simeq -\,\frac{\bigl(p_{\mathrm{m}}+\epsilon_{\mathrm{m}}\bigr)\bigl(m+4\pi r^{3}p\bigr)}
              {r^{2}-2mr} \notag\\
&\quad -\,\frac{\epsilon_{\mathrm{m}}}{\epsilon_{\mathrm{m}}+\epsilon_{\mathrm{de}}}
              \;\frac{\bigl(p_{\chi}+\epsilon_{\chi}\bigr)\bigl(m+4\pi r^{3}p\bigr)}
              {r^{2}-2mr}\,. 
\end{align}
All explicit dependence on $\alpha$ cancels, demonstrating that $dp_{\mathrm{m}}/dr$ is also insensitive to variations in the coupling constant.

In summary, because the three fluid equations self-regulate in this way, the stellar structure responds very weakly to $\alpha$.

\subsection{Results of Model~III: interacting fluids with a unified dark sector}
\label{sec:results_model_3}

\begin{figure}
\centering     
\includegraphics[width=0.48\textwidth]{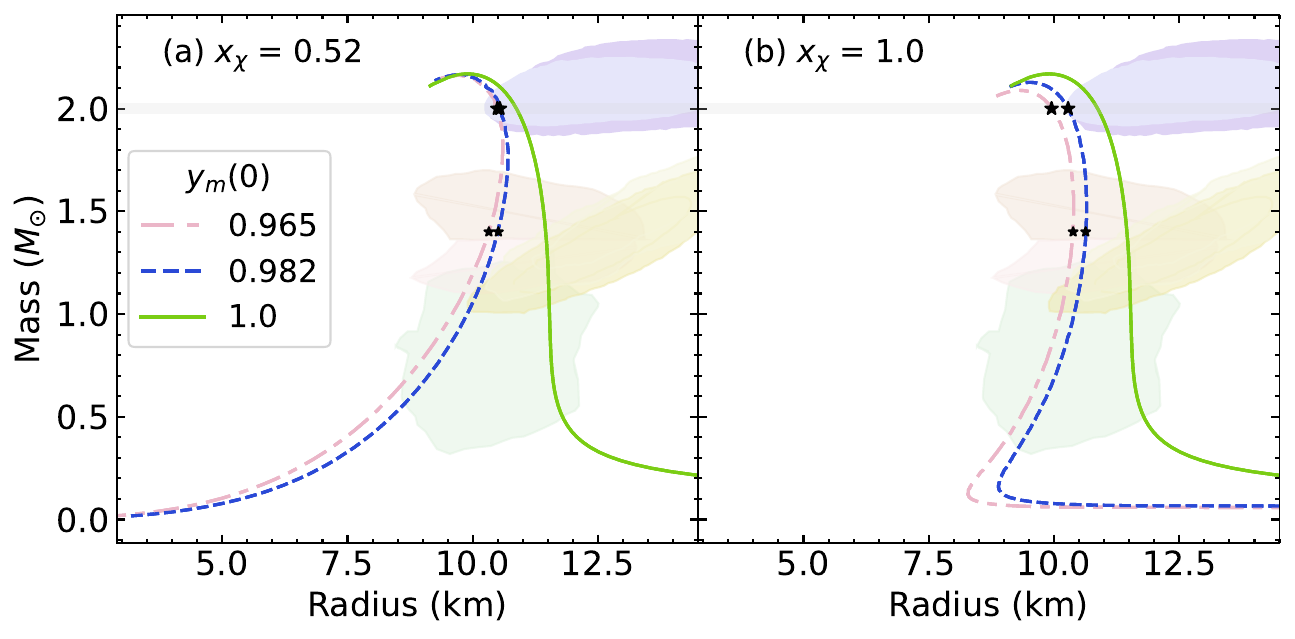}
\caption{Mass--radius relations for Model~III. Dark matter particles are assumed to have a mass of $m_\chi=10\,\mathrm{GeV}$ with no self-interaction, and we set $\alpha=1$ in the interaction term of Eq.~\eqref{eq:Q_definition_model3}.}
\label{fig:8}
\end{figure}

\begin{figure}
\centering     
\includegraphics[width=0.48\textwidth]{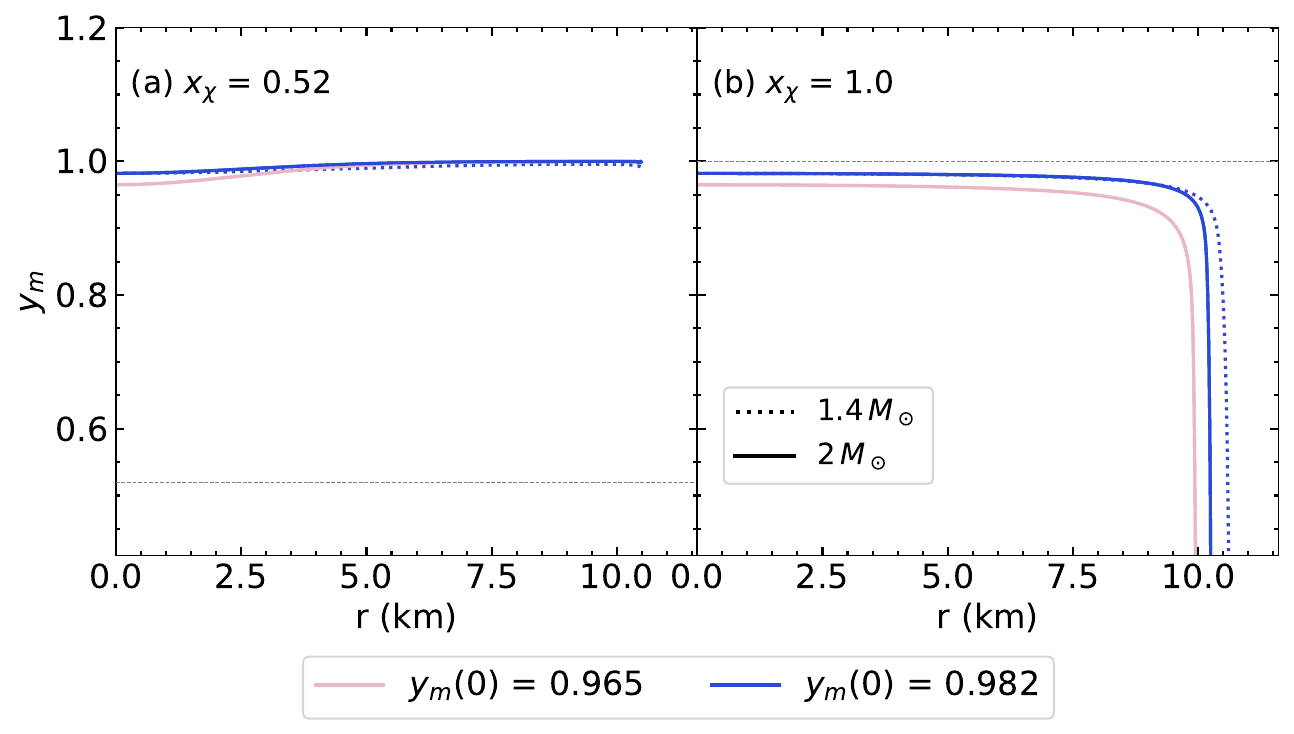}
\caption{Radial profiles of the baryonic fraction $y_{\mathrm{m}}(r)$ and the DM fraction $x_{\chi}$ (horizontal lines) in neutron stars of $1.4\,M_{\odot}$ (canonical) and $2.0\,M_{\odot}$ (high mass) for Model~III. The model parameters are the same as in the previous figure. }
\label{fig:9}
\end{figure}

\begin{figure}
\centering    
\includegraphics[width=0.48\textwidth]{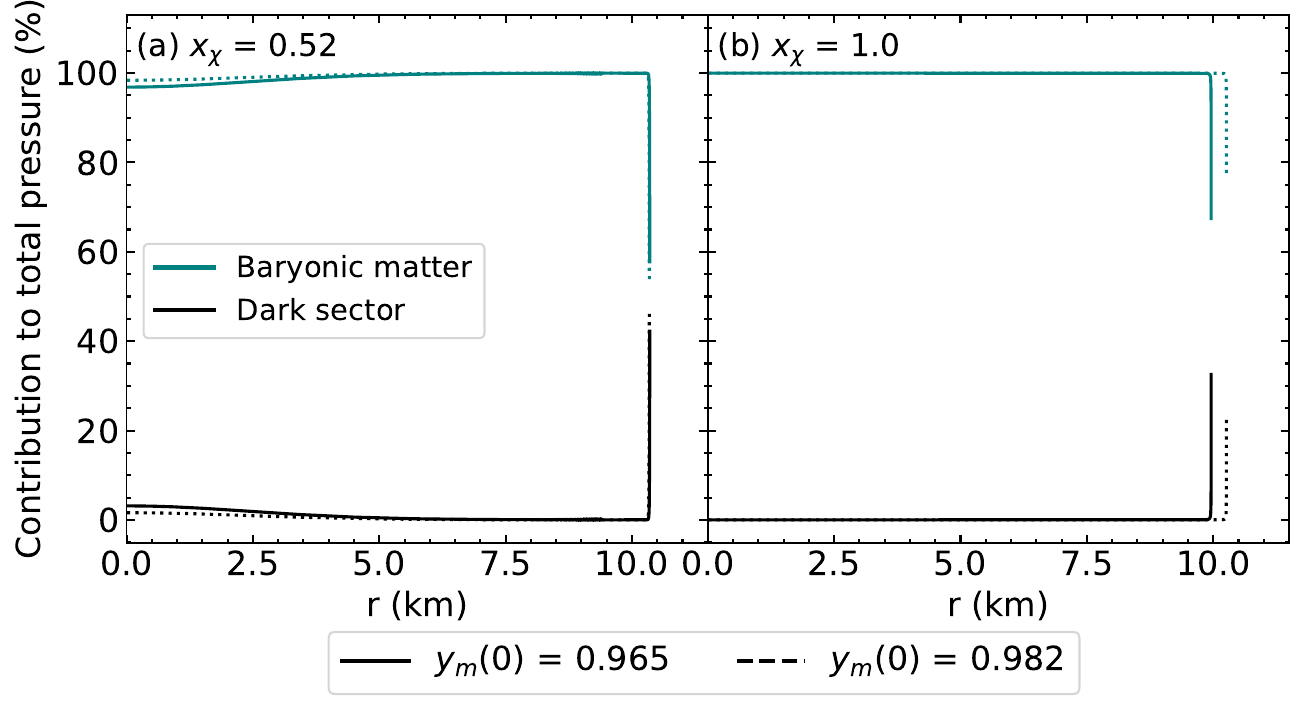}
\caption{Radial profiles for the contribution to the total pressure of each component. The model parameters are the same as in the previous figure.}
\label{fig:10}
\end{figure}

Figures~\ref{fig:8}, \ref{fig:9}, and \ref{fig:10} present the predictions of Model~III, in which DM and DE are merged into a unified dark sector that exchanges energy with baryons via a single interaction term $Q$. In this model, the internal partition between matterlike and vacuumlike contributions is fixed by a constant ratio $x_{\chi}$ at every radial shell. 

Figure~\ref{fig:8}(a) shows the M--R curves when the dark sector is almost equally divided between DM and DE ($x_{\chi} = 0.52$), whereas Fig.~\ref{fig:8}(b) corresponds to a dark sector composed entirely of DM ($x_{\chi} = 1.0$). In both panels, four different baryonic fractions $y_{\mathrm{m}}$ are considered, following the approach used in Models~I and II. As in previous models, lower baryonic fractions yield smaller maximum masses and radii, a trend that is particularly evident for low-mass stars.

Figure~\ref{fig:9} shows how the baryonic fraction $y_{\mathrm m}(r)$ and the DM fraction $x_{\chi}(r)$ vary from the center ($r=0$) to the stellar surface ($r=R$) for two representative neutron star models: a canonical star with $1.4\,M_{\odot}$ and a massive star with $2.0\,M_{\odot}$.  These configurations are marked by asterisks in Fig.~\ref{fig:8}. The trends follow directly from the exchange term $Q=\alpha\,d\epsilon_{\text{dark}}/dr$, which is strictly negative because $\epsilon_{\text{dark}}(r)$ decreases monotonically outward. A negative $Q$ drains pressure and energy from the baryonic fluid more rapidly than would occur in the absence of coupling. In some curves $y_{\mathrm m}(r)$ nevertheless rises slightly in the outer layers; this merely reflects the fact that $\epsilon_{\text{dark}}$ falls off faster than $\epsilon_{\mathrm m}$ in those regions and \emph{does not} imply that the dark‑sector energy density increases with radius.

Figure~\ref{fig:10} illustrates how the total pressure is partitioned radially between baryons and the unified dark sector. With $x_{\chi}=0.52$, the  contribution from DE in the outer layers results in negative pressure effects that render low-mass stars with smaller radii, as apparent from Fig. \ref{fig:8}(a). In contrast, when $x_{\chi}=1.0$, the dark sector does not exert negative pressure, and the resulting pressure distribution is less stiff, with comparatively slightly larger radii.

\begin{figure}[tb]
\centering    
\includegraphics[width= 0.48 \textwidth]{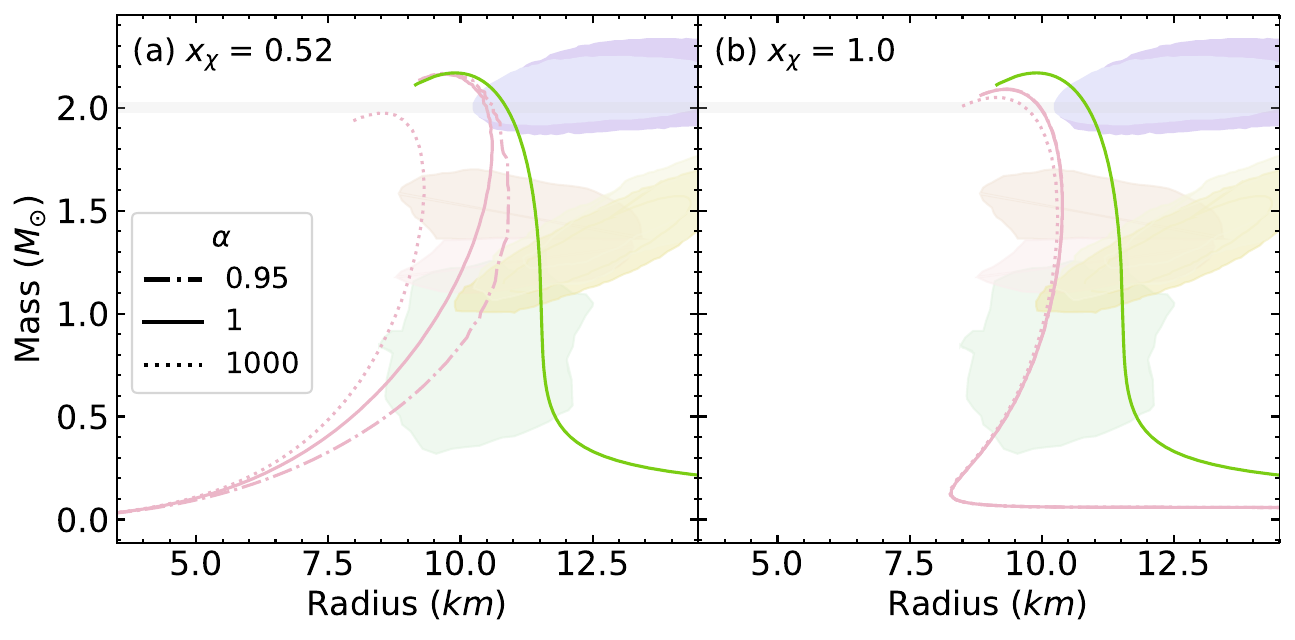}    
\caption{Mass–radius relations for Model~III under the APR EOS, illustrating the sensitivity to the coupling constant $\alpha$ in the unified dark‐sector interaction scenario.  Panel~(a) shows a mixed dark component with $x_{\chi}(0)=0.52$, while panel~(b) corresponds to a purely dark‐matter core with $x_{\chi}(0)=1.0$.   The baryonic fraction $y_{\mathrm m}(0)$ is held fixed at 0.965  and the shaded bands mark observational constraints from pulsar and gravitational‐wave measurements.}
\label{fig:11}
\end{figure}

Figure~\ref{fig:11} examines the influence of the coupling constant $\alpha$ on the stellar sequences predicted by Model~III. In the mixed dark-sector case ($x_{\chi}(0)=0.52$), increasing $\alpha$ from $0.95$ to $1000$ produces a pronounced decrease in both the maximum mass and the radii. Each sequence preserves the self-bound shape induced by the negative-pressure component, but the binding becomes progressively stronger as $\alpha$ increases. By contrast, when the core is purely composed of DM ($x_{\chi}(0)=1.0$), the dependence on $\alpha$ is almost negligible.

To understand this behavior, let us first note that energy exchange in Model~III is governed by the single source term
\begin{equation}
\label{eq:Q_MIII}
Q = \alpha \frac{d\epsilon_{\mathrm{dark}}}{dr} = \alpha \frac{d\epsilon_{\mathrm{dark}}}{dp_{\mathrm{dark}}} \frac{dp_{\mathrm{dark}}}{dr},
\end{equation}
with $\alpha>0$. Combining Eqs.~\eqref{eq:dp_m_unified}–\eqref{eq:Q_definition_model3}, one obtains
\begin{equation}
\label{eq:MIII_dark_grad}
\bigl(1+\alpha\kappa\bigr) \frac{dp_{\mathrm{dark}}}{dr} = -\bigl(p_{\mathrm{dark}}+\epsilon_{\mathrm{dark}}\bigr) H,
\end{equation}
\begin{equation}
\label{eq:MIII_baryon_grad}
\frac{dp_{\mathrm{m}}}{dr} = -\bigl(p_{\mathrm{m}}+\epsilon_{\mathrm{m}}\bigr) H - \frac{\alpha\kappa}{1+\alpha\kappa} \bigl(p_{\mathrm{dark}}+\epsilon_{\mathrm{dark}}\bigr) H,
\end{equation}
where $H(r) = (m+4\pi r^{3}p)/(r^{2}-2mr)$ and $\kappa = d\epsilon_{\mathrm{dark}}/dp_{\mathrm{dark}}$. For the parameters adopted here ($m_\chi = 10\,\mathrm{GeV}$ and $k_{F\chi} \ll m_\chi$),
\begin{equation}
\kappa \simeq \frac{d\epsilon_\chi}{dp_\chi} \approx \frac{5m_\chi^{2}}{k_{F\chi}^{2}} \gtrsim 10^{4},
\end{equation}
so that $\alpha\kappa \gg 1$.

\emph{Dark-sector self-regulation.}  
Equation~\eqref{eq:MIII_dark_grad} then gives  
\begin{equation}
\frac{dp_{\mathrm{dark}}}{dr}
   \simeq -\frac{p_{\mathrm{dark}}+\epsilon_{\mathrm{dark}}}{\alpha\kappa}\,H,
\end{equation}
implying  
\(
Q=\alpha\kappa\,dp_{\mathrm{dark}}/dr
      \simeq -(p_{\mathrm{dark}}+\epsilon_{\mathrm{dark}})\,H,
\)
so the dark sector still self-regulates to the hydrostatic scale.

\emph{Baryonic response and the role of dark energy.}  
Although the algebraic structure of Eq.\,\eqref{eq:MIII_baryon_grad} is identical for $x_{\chi}=0.52$ and $x_{\chi}=1.0$, the physical outcome differs because only the mixed case contains a substantial vacuum component.  With $x_{\chi}=0.52$,  every joule extracted from baryons appears partly as \emph{negative} pressure, so the total pressure $p=p_{\mathrm m}+p_\chi-\epsilon_{\mathrm{de}}$ and hence $H$ decrease appreciably.  The hydrostatic support is therefore weakened, giving the significant softening seen in Fig.~\ref{fig:11}(a).  When $x_{\chi}=1.0$ the dark sector is a cold Fermi gas; the energy transferred from baryons adds almost no pressure, the negative-pressure channel is absent, and the net change in $p$ (and thus in $H$) is small even for $\alpha=1000$.  Consequently, the $M$–$R$ curves in Fig.~\ref{fig:11}(b) shift only slightly with~$\alpha$.

\section{Summary and Conclusions}
\label{sec:conclusions}

We have performed a unified treatment of neutron‐star structure in the presence of three coexisting fluids—baryons, fermionic dark matter with $m_\chi = 10\;\mathrm{GeV}$, and dark energy described by $p_{\mathrm{de}} = \omega\,\epsilon_{\mathrm{de}}$. The distinguishing feature of this work is the explicit inclusion of energy exchange among the fluids, implemented via one or two source terms $Q_i$ that couple local energy–density gradients into the TOV system.  We examined three hierarchical scenarios:

\begin{enumerate}
  \item Model~I:  Baryons, dark matter, and dark energy coexist with fixed fractional decompositions $y_{\mathrm m}$ and $x_\chi$ at each radius, but no explicit energy transfer ($Q_i=0$).
  \item Model~II:   Baryons, DM, and DE exchange energy through two independent source terms $Q_{\mathrm m}$ and $Q_{\mathrm{de}}$, and the DM fluid can also carry a repulsive self‐interaction.
  \item Model~III: DM and DE form a single “dark” fluid that couples to baryons via
  $Q=\alpha\,d\epsilon_\mathrm{dark}/dr$, while the internal DM/DE partition remains fixed by a constant $x_\chi$.
\end{enumerate}

In all three models, the dark sector produces two complementary effects that consistently shape the outcome:

\begin{itemize}
  \item \emph{Softening by massive dark matter}: Heavy fermions ($m_\chi=10\,$GeV) contribute primarily to the energy density but exert negligible pressure, reducing the overall stiffness of the combined EOS and thereby lowering both the maximum mass $M_\mathrm{max}$ and the radius $R$.
  \item \emph{Vacuum softening and binding}: A negative pressure component further softens the EOS in every layer where it is present, by reducing the total pressure.  If this vacuum component extends to the stellar surface, its negative pressure shifts the zero‑pressure boundary to finite density and yields self‑bound configurations reminiscent of strange quark stars.
\end{itemize}

These two mechanisms constitute the dominant drivers of the star’s structure.  In fact, across the three models, it is the baryonic fraction $y_{\mathrm m}$ and the dark‐sector partition $x_{\chi}$ whose variation produces the largest changes in the mass-radius relations.  Adjusting $y_{\mathrm m}$ changes the relative weight of high‐pressure baryons versus low‐ or negative‐pressure dark components, while tuning $x_{\chi}$ alters the balance between massive, pressure‐poor dark matter and vacuum‐like dark energy.  Together, these parameters and the softening and binding effects they control serve as the principal levers governing neutron star compactness, maximum mass, and the emergence of self‐bound configurations.

Having established the dominant roles of DM/DE induced softening and vacuum binding in setting the stellar compactness, we now examine how explicit energy exchange further shapes the structure. 
In Model I (no transfer), the fractional profiles remain fixed, and these mechanisms act in their simplest form. When $x_\chi=0.47$ and $\epsilon_\star=0$ the star becomes self‑bound, whereas a purely DM star ($x_\chi=1$) remains gravitationally bound but clearly softened.  Increasing either $\epsilon_\star$ or the baryon fraction $y_{\mathrm m}$ pushes the solution back toward the purely hadronic sequence.
In Model II, the energy exchange terms exhibit strong self‐regulation so that every $Q_i$ remains locked to the hydrostatic scale.  All explicit $\alpha$–dependence cancels from the leading gradients leaving the global EOS unchanged. Hence, the M--R curves are virtually identical for any choice of the coupling constant $\alpha$. 
In Model III, the self-regulation of the dark sector survives, but $\alpha$ now affects the baryonic gradient.   If the dark sector contains a substantial vacuum fraction ($x_\chi=0.52$) the negative pressure of dark energy amplifies the impact of $Q$: increasing $\alpha$ transfers baryonic energy into a component that adds negative pressure, so the total pressure falls significantly, the EOS softens, the stellar radii decrease, and $M_{\mathrm{max}}$ drops.   In contrast, for a pure DM core ($x_\chi=1$), the added energy enters a cold Fermi gas whose pressure is minuscule; the net change in total pressure is therefore tiny, and the M--R curve shifts slightly even when $\alpha$ increases by three orders of magnitude.   Thus Model~III breaks the $\alpha$-degeneracy found in Model II, but only when the dark sector carries a significant negative-pressure component.

We close by stressing the exploratory character of the present study.  All numerical results were obtained with a deliberately simple microphysical setup: a single fermionic dark-matter species of fixed mass $m_\chi=10~\mathrm{GeV}$, a barotropic dark-energy component with constant $\omega$, and phenomenological couplings $Q_i$ whose strengths were chosen \emph{ad hoc}.  None of these ingredients is compelled by either laboratory data or cosmology, and in several respects (most conspicuously the local dark-energy density, which exceeds the cosmological value by $\gtrsim40$ orders of magnitude) our parameter space is manifestly non-standard.  The aim has therefore not been to predict observable signatures that could diagnose the dark sector of a neutron star—an enterprise that would require a far more tightly constrained microphysics—but rather to map the \emph{qualitative} responses of cold, compact matter to generic forms of dark softening, vacuum binding, and baryon–dark energy exchange.  By working within fully relativistic stellar structure and allowing the three fluids to interact, we have provided a controlled baseline against which future, more realistic models can be gauged.  In that sense, the present framework should be viewed as a theoretical laboratory: it identifies which combinations of dark-sector properties \emph{could} have macroscopic impact, delineates the conditions under which that impact is self-regulated or amplified, and supplies analytic scalings that remain valid whatever microphysical completion is eventually adopted.

\section*{Acknowledgments}
L.F.A. was supported by Coordenação de Aperfeiçoamento de Pessoal de Nível Superior (CAPES), Brazil. J. A. S. L. is partially supported by the Brazilian agencies, CNPq under Grant No. 310038/2019-7, CAPES (88881.068485/2014) and FAPESP (LLAMA Project
No. 11/51676-9). G.L. acknowledges the financial support from CNPq (grant 316844/2021-7).

\bibliography{references}
\end{document}